\documentclass{emulateapj}
\usepackage{amssymb,amsmath}
\usepackage{verbatim}
\usepackage{color,hyperref}
\usepackage{needspace}
\definecolor{linkcolor}{rgb}{0,0,0.25}
\hypersetup{
  colorlinks=true,        
  linkcolor=linkcolor,    
  citecolor=linkcolor,    
  filecolor=linkcolor,    
  urlcolor=linkcolor      
}
\newcounter{address}
\makeatletter
\newcommand{\manuallabel}[2]{\def\@currentlabel{#2}\label{#1}}
\newcommand{\pushright}[1]{\ifmeasuring@#1\else\omit\hfill$\displaystyle#1$\fi\ignorespaces}
\newcommand{\pushleft}[1]{\ifmeasuring@#1\else\omit$\displaystyle#1$\hfill\fi\ignorespaces}
\makeatother
\newcommand{\ie}{i.e.}
\newcommand{\etal}{et al.}
\newcommand{\dd}{\mathrm{d}}
\newcommand{\eg}{e.g.}
\newcommand{\eqnname}{equation}
\newcommand{\Eqnname}{Equation}
\newcommand{\equationname}{\eqnname}
\newcommand{\Equationname}{\Eqnname}

\renewcommand{\figurename}{Figure}

\newcommand{\sectionname}{$\mathsection$}

\newcommand{\feh}{\ensuremath{[\mathrm{Fe/H}]}}

\newcommand{\jk}{\ensuremath{(J-K_s)_0}}
\newcommand{\jksq}{\ensuremath{[J-K_s]_0}} 

\newcommand{\dens}{\ensuremath{\nu_*}}

\newcommand{\kpc}{\ensuremath{\,\mathrm{kpc}}}
\newcommand{\pc}{\ensuremath{\,\mathrm{pc}}}

\newcommand{\magunit}{\,\mbox{mag}}

\newcommand{\field}{\ensuremath{\mathrm{field}}}
\usepackage{yfonts}
\newcommand{\essf}{\ensuremath{\textswab{S}}}

\submitted{}

\begin{document}

\title{On Galactic density modeling in the presence of dust extinction}

\author{Jo~Bovy\altaffilmark{1,2,3},
  Hans-Walter~Rix\altaffilmark{4},
  Gregory~M.~Green\altaffilmark{5},
  Edward F. Schlafly\altaffilmark{4},
  and Douglas P. Finkbeiner\altaffilmark{5}}
\altaffiltext{\theaddress}{\label{1}\stepcounter{address} 
  Department of Astronomy and Astrophysics, University of Toronto, 50
  St.  George Street, Toronto, ON, M5S 3H4, Canada;
  bovy@astro.utoronto.ca~}
\altaffiltext{\theaddress}{\label{2}\stepcounter{address} 
  Institute for Advanced Study, Einstein Drive, Princeton, NJ 08540,
  USA}
\altaffiltext{\theaddress}{\label{3}\stepcounter{address} 
  John Bahcall Fellow}
\altaffiltext{\theaddress}{\label{4}\stepcounter{address} 
  Max-Planck-Institut f\"ur Astronomie, K\"onigstuhl 17, D-69117
  Heidelberg, Germany}
\altaffiltext{\theaddress}{\label{5}\stepcounter{address} 
  Harvard-Smithsonian Center for Astrophysics, 60 Garden Street,
  Cambridge, MA 02138, USA}

\begin{abstract}  
Inferences about the spatial density or phase-space structure of
stellar populations in the Milky Way require a precise determination
of the effective survey volume. The volume observed by surveys such as
\emph{Gaia} or near-infrared spectroscopic surveys, which have good
coverage of the Galactic mid-plane region, is highly complex because
of the abundant small-scale structure in the three-dimensional
interstellar dust extinction. We introduce a novel framework for
analyzing the importance of small-scale structure in the
extinction. This formalism demonstrates that the spatially-complex
effect of extinction on the selection function of a pencil-beam or
contiguous sky survey is equivalent to a low-pass filtering of the
extinction-affected selection function with the smooth density
field. We find that the angular resolution of current 3D extinction
maps is sufficient for analyzing \emph{Gaia} sub-samples of millions
of stars. However, the current distance resolution is inadequate and
needs to be improved by an order of magnitude, especially in the inner
Galaxy. We also present a practical and efficient method for properly
taking the effect of extinction into account in analyses of Galactic
structure through an effective selection function. We illustrate its
use with the selection function of red-clump stars in APOGEE using and
comparing a variety of current 3D extinction maps.
\end{abstract}


\section{Introduction}

\begin{figure*}[t!]
\begin{center}
\includegraphics[width=0.99\textwidth,clip=]{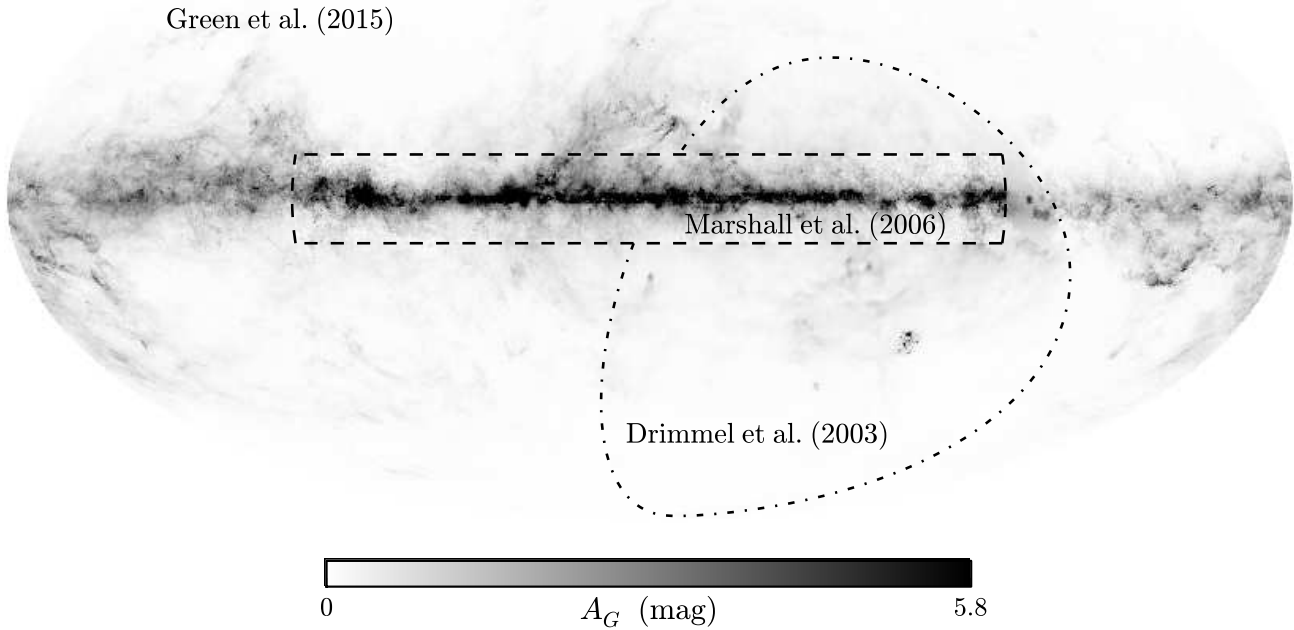}
\end{center}
\caption{$G$-band extinction to $5\kpc$ from the Sun, saturated at the
  approximate maximum $A_G$ for which we can see the RC at $G < 20$
  (that is, in \emph{Gaia}). For the APOGEE $H$-band observations,
  this extinction map saturates at $A_H \approx1.1$, which is the
  maximum extinction to which the RC can be seen to $\approx4.5\kpc$
  in medium-deep APOGEE observations. The boundaries of the three
  extinction maps that are combined to obtain full-sky coverage (see
  the \appendixname) are indicated. Much of the Galactic plane region
  is too extinguished for \emph{Gaia} or APOGEE to be able to see the
  RC at $\approx5\kpc$ and larger distances.}\label{fig:dust}
\end{figure*}

A common problem in the analysis of stellar surveys in the Milky Way
is to account for the effects of astrophysical and procedural
selection effects on the manner in which the Galactic volume is
surveyed \citep[\eg,][]{Rix13a}. The latter effects primarily consist
of the fraction of stars observed as a function of their position on
the celestial sphere, stellar type, color, or magnitude. Examples of
the astrophysical effects include how interstellar extinction impacts
the observed magnitudes of stars or the manner in which different
stellar types trace the underlying stellar populations. Accounting for
these selection biases is essential for using stellar surveys to
determine the spatial or phase--space distribution of stars in our
Galaxy.

The basic formalism for fitting models of the density or phase--space
distribution has been reviewed by \citet{Rix13a}. The formalism
presented there is general, but in this paper we focus on
understanding the formalism and its implementation in the case that
the effect of interstellar extinction on the selection function of a
stellar survey is significant. To motivate this,
\figurename~\ref{fig:dust} displays the amount of interstellar
extinction to $5\kpc$ over the whole sky from a combination of the
\citet{Marshall06a}, \citet{Green15a}, and \citet{Drimmel03a}
extinction maps (see the \appendixname). It is clear from this map
that any investigation of the phase--space distribution of the stellar
disk near the midplane is strongly affected by the highly filamentary
structure of the extinction. Determining simple cuts to define a
volume-complete sample of stars in this region of the Galaxy is
difficult and sub-optimal.

Below, we present a simple and efficient formalism for determining and
understanding the effect of extinction on the effective volume of a
survey. We do this for two cases: (a) a large-area survey such as
\emph{Gaia} that covers a significant fraction of the sky and (b) a
pencil-beam survey, such as the spectroscopic APOGEE survey
(S.~R.~M. Majewski, \etal, 2015, in preparation), consisting of a
number of small-area field pointings. While the formalism is different
in practice, in both cases an important conclusion is that the effect
of extinction on small spatial scales is unimportant when considering
smooth models for the density or phase--space distribution of stars.

The outline of this paper is as follows. \sectionname~\ref{sec:like}
reviews the basic likelihood-based formalism for inferring stellar
phase--space distributions. We present the formalism for determining
the effective survey volume necessary in the likelihood for large-area
surveys and pencil-beam surveys in
\sectionname\sectionname~\ref{sec:large-formalism} and
\ref{sec:pencil-formalism}, respectively. We give a detailed example
of a \emph{Gaia}-like large-area survey of red-clump (RC) stars in
\sectionname~\ref{sec:large-example}. \sectionname~\ref{sec:pencil-example}
discusses the effect of extinction on the pencil-beam APOGEE volume
selection of RC stars as an example of the formalism in
\sectionname~\ref{sec:pencil-formalism}. We discuss and conclude in
\sectionname~\ref{sec:discussion}.

\section{Likelihood-based density modeling and the effective survey volume}\label{sec:like}

As discussed in detail by \citet{BovyMAPstructure} and \citet{Rix13a},
the correct model of star counts is an inhomogeneous Poisson
process. This process is characterized by a rate function
$\lambda(O|\theta)$ that specifies the model prediction for the number
of stars as a function of $O = (l,b,D,m,c,\feh)$; the model is
parameterized by parameters $\theta$. Here we have written the
observables in a generic form consisting of: (a) the position in
Galactic longitude, latitude, and heliocentric distance $(l,b,D)$, (b)
an apparent magnitude $m$, and (c) a color $c$ and metallicity
\feh. It is necessary to include the distribution of color (or
multiple colors) and potentially metallicity when the absolute
magnitude $M$ of the stellar tracer being considered depends on
these. We will see below that the formalism is especially simple in
the case of a standard candle, that is, a stellar tracer for which $M$
is constant; however it is straightforward to include the full
dependence of $M$ on color(s) and metallicity, or any other observable
that affects the absolute magnitude.

The rate function $\lambda(O|\theta)$ is given by
\begin{equation}
\begin{split}
  \lambda(O|\theta) & = \dens(X,Y,Z|\theta)\times|J(X,Y,Z;l,b,D)|\\
  & \ \times\rho(M,c,\feh|X,Y,Z)\times S(l,b,m)\,,
\end{split}
\end{equation}
where $\dens(\cdot|\theta)$ is the spatial density in Galactocentric
rectangular coordinates $(X,Y,Z)$ that we are ultimately most
interested in and that depends on parameters $\theta$,
$|J(X,Y,Z;l,b,D)| = D^2\,\cos b$ is the Jacobian of the transformation
between $(X,Y,Z)$ and $(l,b,D)$, $\rho(M,c,\feh|X,Y,Z)$ is the density
of stars in absolute-magnitude--color--metallicity space (normalized
to integrate to one), and $S(l,b,m)$ is the survey selection function
(the fraction of stars from the underlying population of potential
targets observed by the survey). For simplicity we have assumed that
the survey selection function does not depend on color, metallicity,
or any other observable, but including these is straightforward.

As in \citet{BovyMAPstructure}, the rate has an additional amplitude
parameter that gives the number density of stars at a reference
location. The full log likelihood for the Poisson process is
\begin{equation}
  \ln \mathcal{L}(\theta) = \sum_i\ln \lambda(O_i|\theta) -
    \int \dd O \lambda(O|\theta)\,.
\end{equation}
The integral in this equation is the expected number of stars given
the model parameters that provides the normalization of the rate
likelihood. It does not depend on the individual data point, but is
instead a property of the whole survey for a given model specified by
$\theta$. In what follows we refer to this integral as the
\emph{effective volume $V_\mathrm{eff}$ of the survey}, because it is
proportional (up to a constant, the inverse local density) to the
traditional definition of the effective volume.

In many cases, the overall amplitude parameter is uninteresting. To
remove the amplitude parameter from further consideration, we can
marginalize the probability of the parameters of the rate function
over the amplitude of the rate. Using a prior on the amplitude that is
a power-law with exponent $\alpha$, the marginalized likelihood is
given by
\begin{equation}
  \ln \mathcal{L}(\theta) = \sum_i\ln \left[\frac{\lambda(O_i|\theta)}{V_{\mathrm{eff}}}\right]-(1+\alpha)\,\ln V_{\mathrm{eff}}\,.
\end{equation}
For a prior that is inversely proportional to the amplitude ($\alpha =
-1$), the second term in this equation is zero. The 
likelihood can be simplified to
\begin{equation}\label{eq:like}
  \ln \mathcal{L}(\theta) = \sum_i\ln \left[\frac{\dens(X_i,Y_i,Z_i|\theta)}{V_{\mathrm{eff}}}\right]-(1+\alpha)\,\ln V_{\mathrm{eff}}\,.
\end{equation}
when the rate $\lambda(O_i|\theta)$ only depends on $\theta$ through
$\dens(\cdot|\theta)$, which is often a good assumption. Then
$\dens(X_i,Y_i,Z_i|\theta)$ is the only factor in
$\lambda(O_i|\theta)$ that depends on $\theta$; the other factors can
be dropped.

It is the effective volume that is difficult to compute in the
presence of filamentary extinction and it is the focus of the
remainder of this paper.

\section{The effective survey volume for large-area surveys}\label{sec:large}

\needspace{8ex}
\subsection{Formalism}\label{sec:large-formalism}

In this section, we work out the formalism for calculating the
effective survey volume in the presence of dust for a contiguous
survey of a large part of the sky. This is useful for analyzing the
spatial structure of the Milky Way from photometric surveys or for
analyzing the full phase--space structure from \emph{Gaia} data. We
assume that the conversion between distance and absolute magnitude can
be written as a function of an unreddened color and metallicity,
although for a photometric survey the latter might be replaced by a
second color or dropped entirely. We also assume that the selection
function is only a function of magnitude---we denote this magnitude in
this section by $G$ with \emph{Gaia} in mind---although the
generalization to include color dependence (whether corrected for
extinction or not) is straightforward. For contiguous surveys the
selection function typically does not explicitly depend on sky
position and we write it as $S(G[\cdot])$, although it could be easily
incorporated in what follows\footnote{The boundaries of a partial-sky,
  contiguous survey can be incorporated by restricting the integration
  area in $(l,b)$ below to that of the survey, although sharp survey
  boundaries should be apodized to avoid ringing in the basis-function
  expansion below}; we use the argument $G[\cdot]$ to indicate the
dependence of $G$ on $(l,b,D,c,\feh)$, but we do not always write
these variables explicitly to keep the expressions cleaner.  We first
write out the integration over $O$ in the effective volume in
\equationname~(\ref{eq:like}) explicitly
\begin{equation}
  \begin{split}
  \int \dd O\,\lambda(O|\theta) = & \iiint \dd l\,\dd b\,\dd D\,D^2\,\cos b\,\Bigg[\nu_*(X,Y,Z|\theta)\\
  & \ \times\,\iint \dd c\,\dd\feh\, \rho(c,\feh)\,S(G[\cdot])\Bigg]\,,
\end{split}
\end{equation}
where we have used that $|J(X,Y,Z;l,b,D)| = D^2\,\cos b$. The integral
over $(c,\feh)$ is the integration over the absolute-magnitude
distribution and it can be efficiently computed using Monte Carlo
integration (see below). We can then write the density
$\dens(X,Y,Z|\theta)$\footnote{When analyzing the full six-dimensional
  phase--space distribution this density is replaced by the
  integration of the 6D DF over velocity, but the formalism given here
  still applies when the selection function is independent of
  kinematics (which it typically is).} as $\dens(l,b,D|\theta)$ and
write its dependence on $(l,b)$ using an orthonormal basis expansion,
say in terms of spherical harmonics $Y^m_\ell(\pi/2-b,l)$
\begin{equation}\label{eq:dens_ylm}
  \dens(l,b,D|\theta) = \sum_{\ell=0}^\infty\,\sum_{m=-\ell}^\ell \nu_{*,\ell m}(D|\theta)\,Y_\ell^m(\pi/2-b,l)\,.
\end{equation}
Then the effective volume becomes
\begin{widetext}
\begin{equation}
  \int \dd O\,\lambda(O|\theta) = \sum_{\ell=0}^\infty\,\sum_{m=-\ell}^\ell \int \dd D  \,D^2\,\nu_{*,\ell m}(D|\theta)
\,\times\,\iint \dd c\,\dd\feh\, \rho(c,\feh)\ \times\iint \dd l\,\dd b\,\cos b\, Y_\ell^m(\pi/2-b,l)\, S(G[\cdot])\,,
\end{equation}
\end{widetext}
(where all of the integrations are nested, not independent) and we can
introduce the \emph{effective selection function $\essf_{\ell m}$
  of order $(\ell,m)$} \refstepcounter{equation}
\begin{equation}\manuallabel{eq:effsel_cont}{\theequation}
\begin{split}
   \essf_{\ell m}(D) = & \iint \dd c\,\dd\feh\, \rho(c,\feh)\\ & \ \times\iint \dd l\,\dd b\,\cos b\,Y_\ell^m(\pi/2-b,l)\,S(G[\cdot])\,.\notag\\&\pushright{(\mathrm{general\ case};\theequation)}
\end{split}
\end{equation}
If $\rho(c,\feh)$ depends on $(l,b,D)$, then that dependence can be
incorporated by moving the integration over $\rho(c,\feh|l,b,D)$ into
the integral over $(l,b)$. For a standard candle,
\equationname~(\ref{eq:effsel_cont}) simplifies to
\refstepcounter{equation}
\begin{align}\begin{split}
  \essf_{\ell m}(D) & = \iint \dd l\,\dd b\,\cos
  b\,Y_\ell^m(\pi/2-b,l)\,S(G[l,b,D])\,.\notag\\&\pushright{(\mathrm{standard\ candle};\theequation)}
\end{split}\end{align}
 The effective volume in terms of $\essf_{\ell m}$ becomes
\begin{align}\label{eq:effvolcont}
  \int \dd O\,\lambda(O|\theta) = \int \dd D \,D^2\,\sum_{\ell=0}^\infty\,\sum_{m=-\ell}^\ell \nu_{*,\ell m}(D|\theta)
\,\essf_{\ell m}(D)\,.
\end{align}

\Equationname~(\ref{eq:effvolcont}) makes it clear that the effective
survey volume is essentially the cross-correlation of the density with
the effective selection function. Because the Galactic stellar
phase-space density is much smoother than the extinction map, this
cross correlation is essentially a low-pass filtering of the
extinction-affected selection function with the density that limits
the influence of the small-scale power in the interstellar dust
distribution.

We discuss the precision to which the effective survey volume can be
computed as a function of the angular and distance resolution in the
3D extinction map below. The required angular resolution depends on
the sample size, the complexity of the model, the sky coverage,
etc. and it can be determined from an analysis of the cross
correlation in \equationname~(\ref{eq:effvolcont}) as in the example
below. Once a good angular resolution is determined one will in
practice prefer to work with the effective selection function in
position space. For this we define the \emph{effective selection function $\essf(l,b,D)$}
\refstepcounter{equation}
\begin{equation}\begin{split}
    \essf(l,b,D) & = \iint \dd c\,\dd\feh\, \rho(c,\feh)\,S(G[\cdot])\,,\notag\\&\pushright{(\mathrm{general\ case};\theequation)}
\end{split}\end{equation}
which can be efficiently computed using Monte-Carlo integration. For a
standard candle we simply have that $\essf(l,b,D) = S(G[\cdot])$.  In
both cases the $(l,b)$ dependence of $\essf(l,b,D)$ is through the
$(l,b)$ dependence of the extinction that gives $G = M_G + \mu +
A_G(l,b,D)$, where $M_G$ is the absolute magnitude and $\mu$ is the
distance modulus. The effective volume is then simply
\begin{equation}\label{eq:effvol-cont-pos}\begin{split}
  \int \dd O\,&\lambda(O|\theta) = \\
  & \iiint \dd l\,\dd b\,\dd D\,D^2\,\cos b\,\nu_*(X,Y,Z|\theta)\,\essf(l,b,D)\,.
\end{split}\end{equation}

Thus, in practice we do not need to work in Fourier space, but the
Fourier space analysis is helpful to elucidate the importance of
small-scale power in the extinction map and to determine an
appropriate resolution to discretize the integration in
\equationname~(\ref{eq:effvol-cont-pos}).

\begin{figure}[t!]
\begin{center}
\includegraphics[width=0.48\textwidth,clip=]{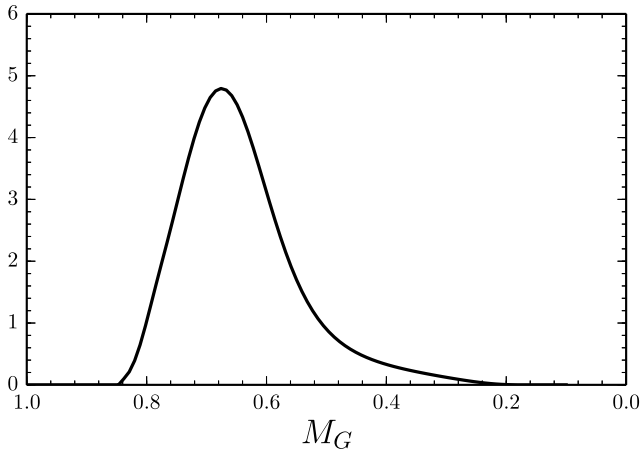}
\end{center}
\caption{Distribution of the absolute $G$-band magnitude derived from
  PARSEC isochrones for an RC sample selected using the cuts from
  \citet{BovyRC} and assuming a solar-neighborhood metallicity
  distribution. The distribution of $M_G$ peaks at $M_G = 0.68$ with an approximate width of $0.08\magunit$.}\label{fig:gaia_mg}
\end{figure}

\begin{figure*}[t!]
\begin{center}
\includegraphics[width=0.47\textwidth,clip=]{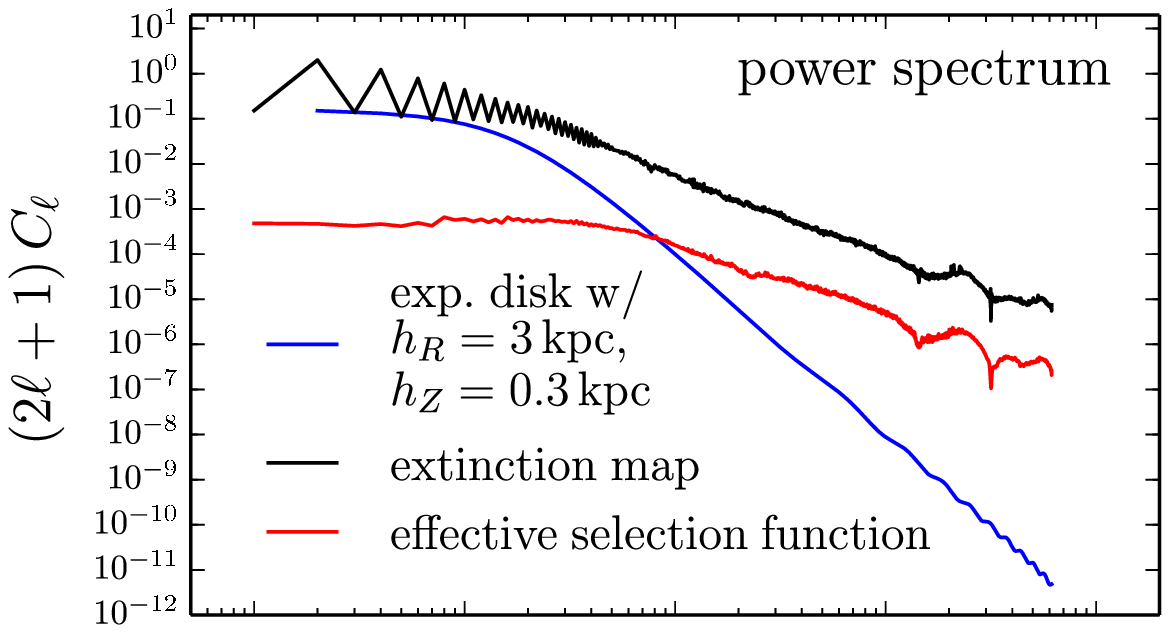}
\includegraphics[width=0.47\textwidth,clip=]{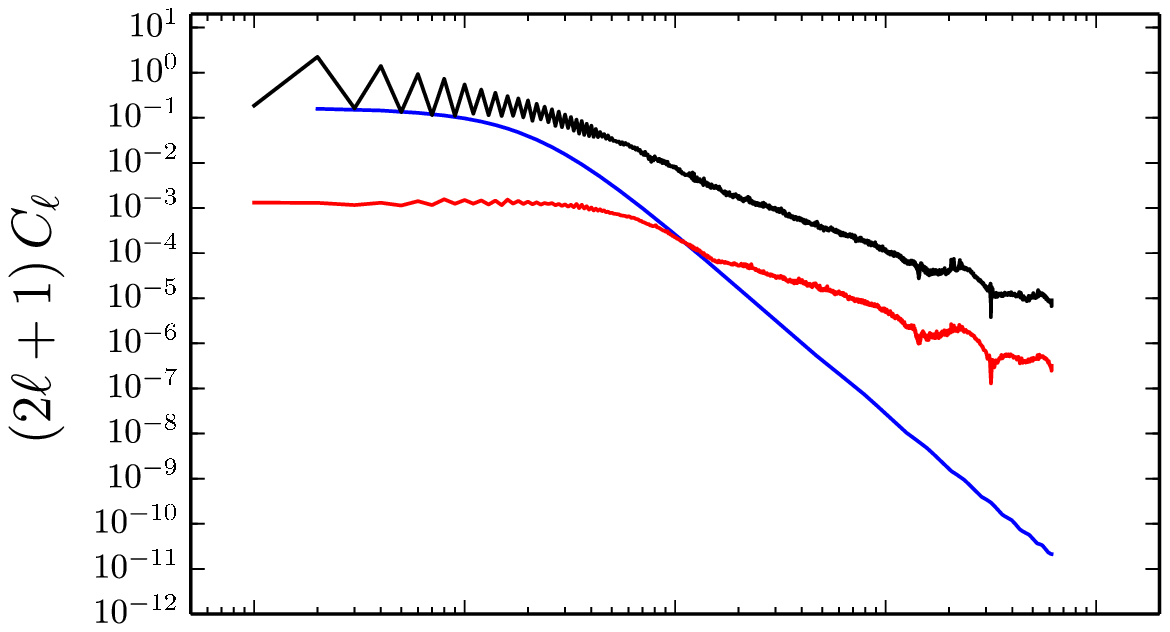}\\
\includegraphics[width=0.47\textwidth,clip=]{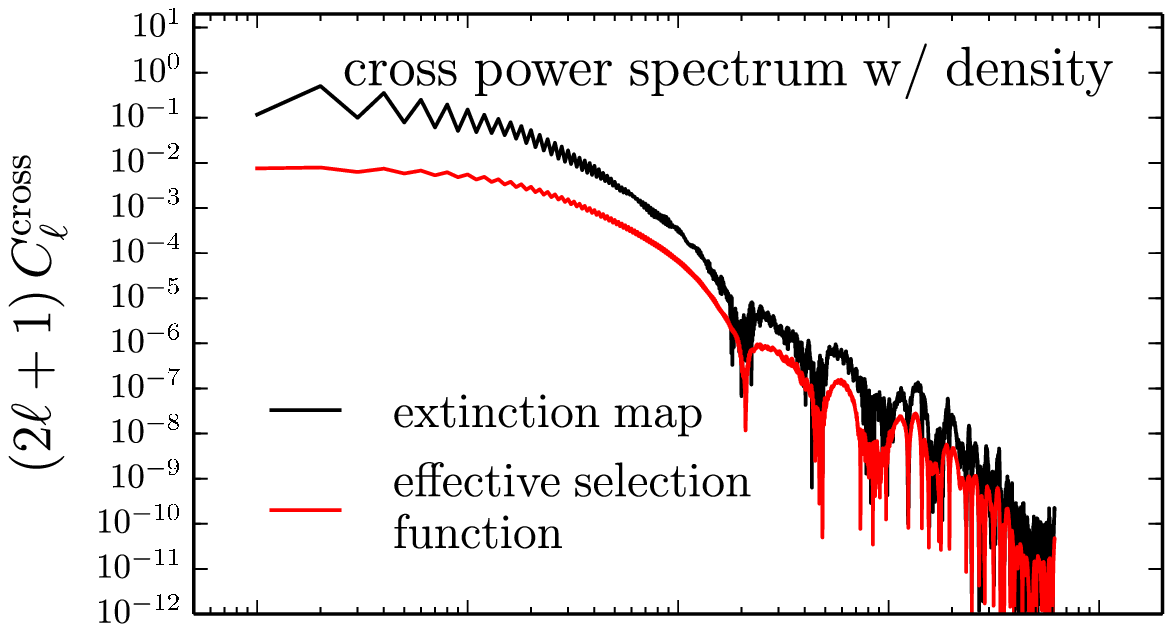}
\includegraphics[width=0.47\textwidth,clip=]{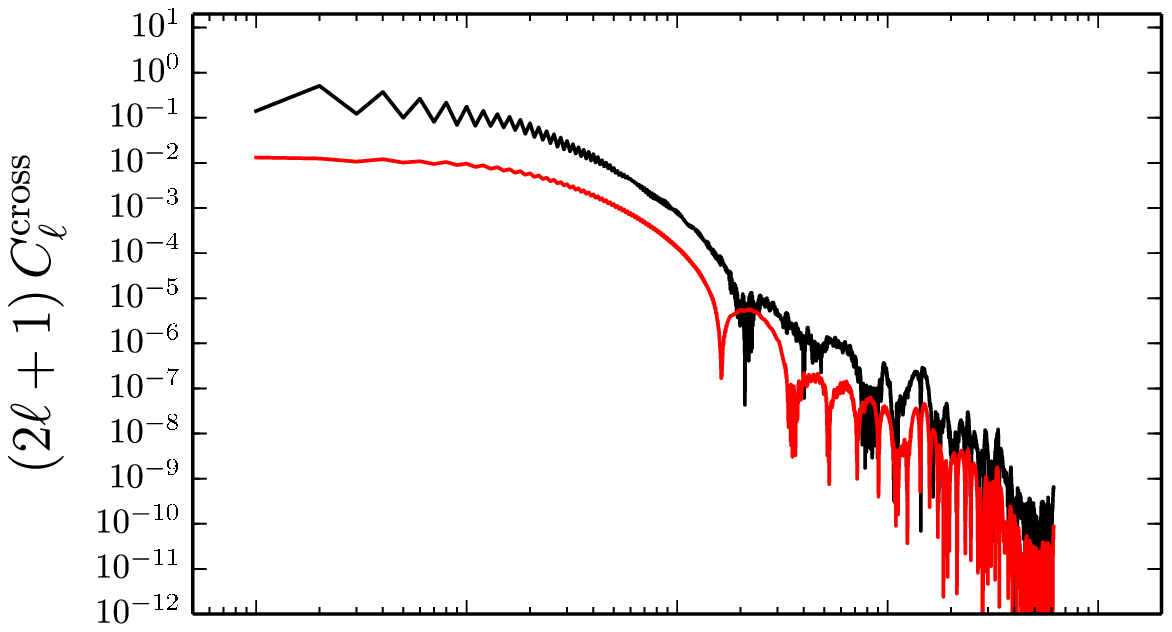}\\
\includegraphics[width=0.47\textwidth,clip=]{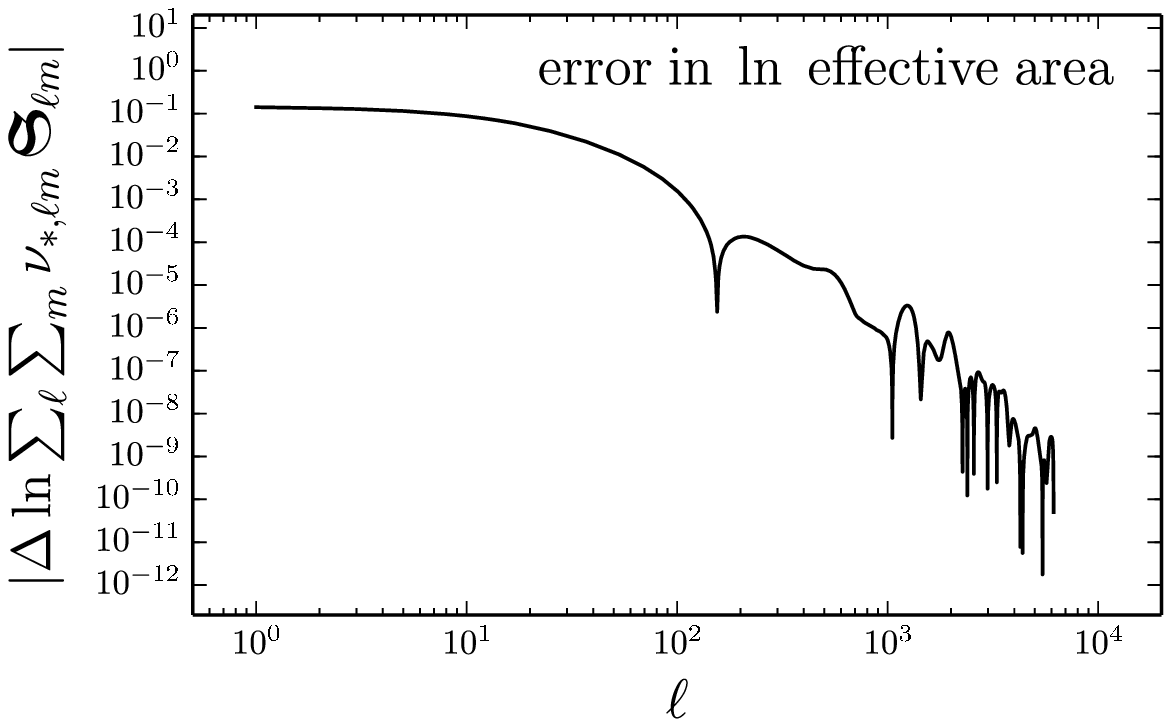}
\includegraphics[width=0.47\textwidth,clip=]{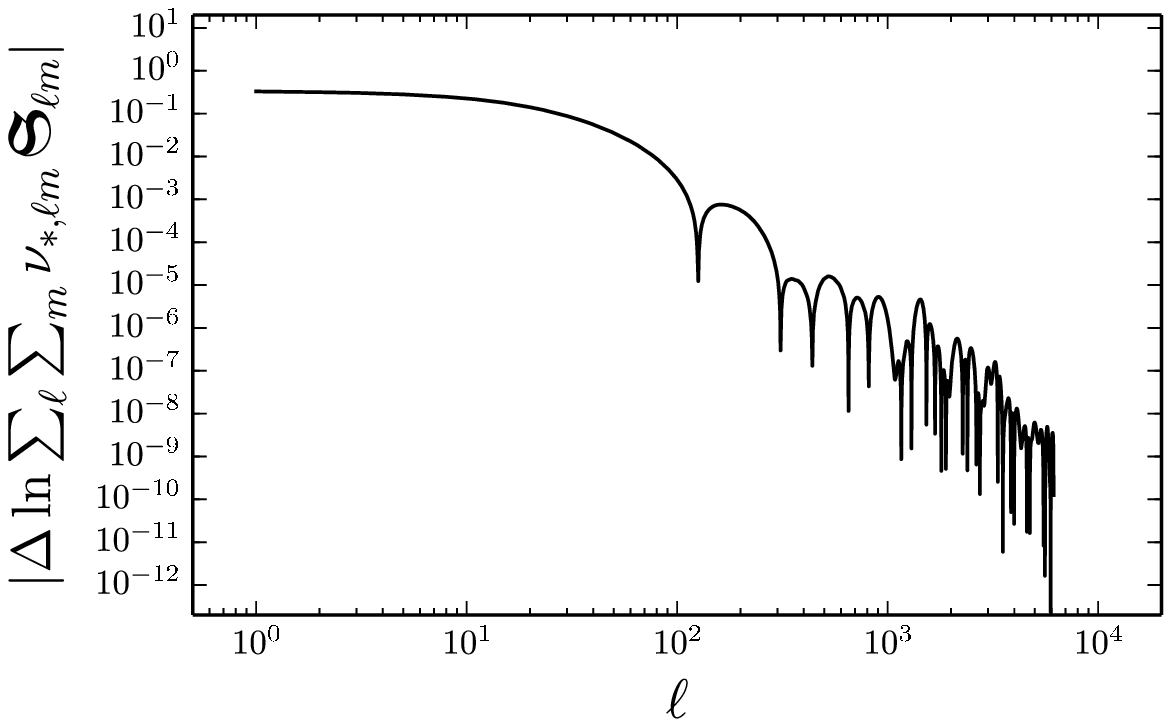}
\end{center}
\caption{Top panels: Power spectrum of an exponential disk (blue; $h_R
  = 3\kpc$, $h_Z = 0.3\kpc$), the combined extinction map (see the
  \appendixname) (black), and the selection function of a
  \emph{Gaia}-like RC sample ($ 3 \leq G \leq 20$; red) at $5\kpc$
  (left panels) and $6.3\kpc$ (right panels). Middle panels: Cross
  power-spectra of the density with the extinction map (black) and the
  selection function (red). Bottom panels: error in calculating the
  natural logarithm of the effective area when only including terms up
  to spherical degree $\ell$ in the calculation: the effective area is
  the cross-correlation of the density and the selection function. The
  pixelization of the extinction maps affects the power spectra at $\ell
  \gtrsim 2\times10^3$. The density acts as a low-pass filter on the
  selection function, whose integrated power would otherwise only
  slowly converge when going to smaller scales.}\label{fig:powspec}
\end{figure*}

\subsection{\emph{Gaia} example}\label{sec:large-example}

\begin{figure}[t!]
  \includegraphics[width=0.46\textwidth,clip=]{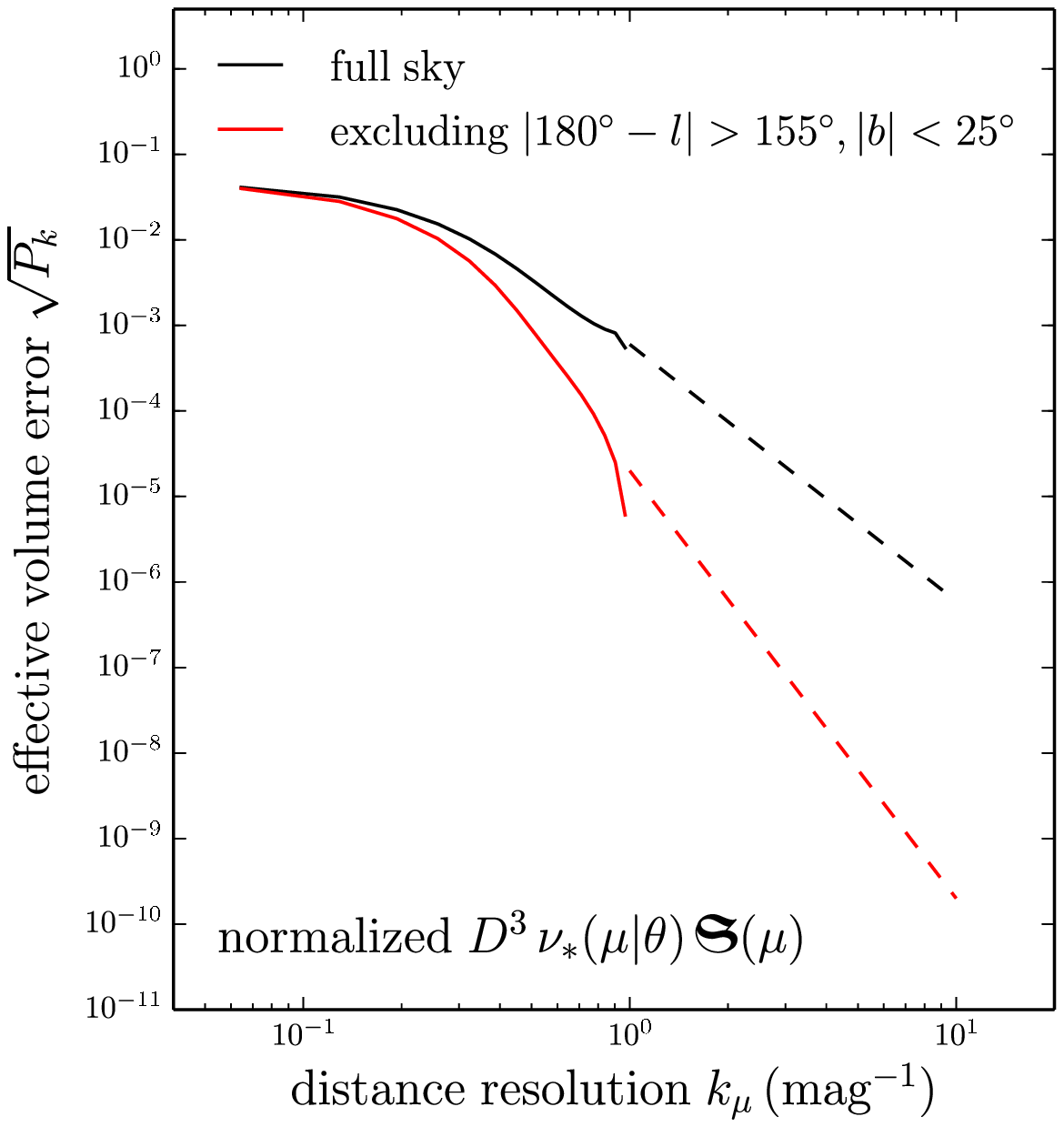}
\caption{Power spectrum of the effective area as a function of
  distance modulus, computed for an exponential disk ($h_R = 3\kpc$,
  $h_Z = 0.3\kpc$) and the selection function of a \emph{Gaia}-like RC
  sample ($ 3 \leq G \leq 20$). This power spectrum can be interpreted
  as the error made in the computation of the effective volume as a
  function of the limited distance sampling of 3D extinction maps. The
  black curve displays the full-sky effective area and the red curve
  excludes a rectangular inner $25^\circ\times25^\circ$ box. The error
  in the effective volume due to the poor distance sampling of
  current extinction maps is $\approx10^{-3}$ over the full sky, but
  dominated by the highly-extinguished inner Galaxy; excluding the
  central region, currently the error is
  $\approx10^{-5}$. Extrapolating the trends with resolution (dashed
  lines), an order of magnitude improvement in the distance resolution
  should lead to errors $\lesssim10^{-6}$ over the full sky and
  $\lesssim10^{-10}$ excluding the central region, which is good
  enough for large-scale \emph{Gaia} analyses.}\label{fig:gaia_dist}
\end{figure}

In this section we work through an example of the formalism described
above to determine the impact of interstellar extinction on
phase-space inferences from \emph{Gaia}. We assume that the stellar
tracers for which we are fitting the density distribution are RC
stars, which are close to the standard candle case. For the purpose of
this example, we estimate the absolute magnitude distribution of RC
stars in the \emph{Gaia} $G$ band as follows. We use PARSEC isochrones
\citep{Bressan12a} and the transformation between $(g,g-z)$ and $G$
from \citet{Jordi10a} to determine the mean absolute magnitude of
stars of a given age and overall metallicity $\feh$ that satisfy the
RC selection criteria of \citet{BovyRC}. We find that the spread at a
given age and metallicity is small and furthermore that the age
dependence is small. Therefore, we determine $M_G(\feh)$ as the mean
$M_G$ at an age of 5\,Gyr (thus effectively ignoring the color
dependence in the integrals above). We use the local metallicity
distribution \citep{Hayden15a} to determine the distribution of $M_G$
from this. The resulting distribution of $M_G$ is displayed in
\figurename~\ref{fig:gaia_mg}. The distribution peaks at $M_G = 0.68$
and is narrow with an approximate width of $0.08\magunit$. We stress
that this is a simplified treatment of the absolute magnitude
distribution, which will in detail depend on the position in the
Galaxy through its dependence on age and metallicity. This dependence
can be taken into account using the formalism above. We use 10,000
Monte Carlo samples from the distribution in
\figurename~\ref{fig:gaia_mg} to approximate the integration over the
intrinsic absolute magnitude distribution above.

We consider a \emph{Gaia}-like survey of RC stars, with a selection
function that is constant over the full sky between $G=3$ and $G=20$
and compute the effective volume for a density model that consists of
a radial and vertical exponential disk with a scale length of $3\kpc$
and a scale height of $300\pc$. The ingredients of the effective
area---the angular part of the integration in
\equationname~(\ref{eq:effvolcont})---at two different distances are
shown in \figurename~\ref{fig:powspec}. The top panels display the
angular power spectra of all of the ingredients: the density
distribution, the extinction distribution, and the effective selection
function of \equationname~(\ref{eq:effsel_cont})\footnote{The $C_\ell$
  and $C^{\mathrm{cross}}_\ell$are computed using HEALPix
  \citep{Gorski05a} using the standard definition \[C_\ell =
  \frac{1}{2\ell+1}\,\sum_m |a_{\ell m}|^2\,,\]
  and \[C^{\mathrm{cross}}_\ell = \frac{1}{2\ell+1}\,\sum_m a_{\ell
    m}\,b^*_{\ell m}\,,\] where $a_{\ell m}$ and $b_{\ell m}$ are the
  coefficients of the decomposition into spherical harmonics of two
  maps. We do this by evaluating the maps at the highest resolution of
  the \citet{Green15a} map ($N_{\mathrm{side}} = 2048$). The
  pixelation affects the power spectra at the largest $\ell \gtrsim
  2\times10^3$; we do not attempt to correct for this here.}. While
most of the power in the extinction map is on large scales, the power
on smaller scales only slowly diminishes. The red line that shows the
angular power spectrum of the effective selection function
demonstrates that most of the power in the extinction map is
transferred to $\essf$. However, the density of a typical exponential
disk has a much stronger decline in power toward small scales. The
middle panels display the cross power spectrum between the extinction
map and the density as well as that between $\essf$ and the
density. As expected, on large scales where the density power spectrum
is close to constant, the cross power spectrum follows that of the
extinction or \essf, but on small scales the power is significantly
reduced due to the lack of small-scale power in the density. The
bottom panels of \figurename~\ref{fig:powspec} show the impact of the
resolution of the extinction map when computing the effective survey
volume. These panels display the error in computing the effective
survey volume when only including terms up to spherical degree $\ell$
in the calculation, that is, when limiting the resolution to
$\approx180^\circ/\ell$.

When performing a likelihood-based inference we need to be able to
compute the relative log likelihood of models within $-2\,\Delta \ln
\mathcal{L} = \Delta \chi^2 \approx \mathcal{O}(1)$ of each other to a
precision that is much smaller than 1. This is such that errors in the
likelihood computation do not affect the uncertainty assigned to model
parameters (which are roughly assigned based on $\Delta \chi^2 \approx
\mathcal{O}(1)$). Because the effective volume normalizes the
probability for each data point (see \equationname~[\ref{eq:like}]),
this requires us to compute the logarithm of the effective volume of
models within $\Delta \chi^2 \approx \mathcal{O}(1)$ to a precision
much better than $1/N$ for a sample of $N$ data points. The bottom
panels of \figurename~\ref{fig:powspec} display the \emph{absolute}
precision in the effective volume (that is, the precision of a single
effective-volume calculation); if this is smaller than $1/N$, then the
relative precision---the precision of the logarithmic difference
between the effective volumes of two similar models---will also be
computed precisely enough. For a \emph{Gaia}-sized survey of RC stars,
$N \approx 10^6$, which requires an extinction map at a resolution of
$\lesssim 10'$ ($\ell \approx 1,000$). This is close to the resolution
in 3D extinction maps that exists in current maps (see the discussion
in the \appendixname). Extrapolating the behavior at $\ell \lesssim
2\times10^3$---where it is not affected by the pixelization of current
extinction maps---in the bottom panels of
\figurename~\ref{fig:powspec}, we find that the full \emph{Gaia} data
set ($N\approx10^9$) requires a resolution $\lesssim2'$, which is at
the limit of current extinction maps. Modest improvements in the
currently available 3D extinction maps will therefore suffice for
computing the effective area for any smooth model for the phase-space
distribution of stars constrained using \emph{Gaia} data. Analyses
using samples with small sizes do not gain from using a
high-angular-resolution effective selection function; an analysis such
as that presented in the bottom panels of
\figurename~\ref{fig:powspec} can be used to determine a good
resolution to degrade the effective selection function to, to minimize
the computational cost for computing the effective selection function.

Computing the effective volume requires us to integrate the effective
area over distance (see \equationname~[\ref{eq:effvolcont}]). While
the angular resolution of 3D extinction maps is high, the distance
resolution of these maps is poor due to, for example, uncertainties in
the distances to the stellar tracers used to determine the
extinction. The spatial sampling in distance modulus of the
\citet{Green15a} and \citet{Marshall06a} maps is $\approx0.5\magunit$,
although we note that this is undersampled by a factor of $\approx2$,
in that a finer distance sampling is possible for the same data. This
resolution is much worse than the angular resolution of these maps
($\approx10'$--$15'$; see the \appendixname), which corresponds to a
relative distance resolution of less than $1\,\%$. However, because
extinction along a line of sight is a cumulative quantity, the
behavior of extinction with distance is typically smoother than the
angular dependence. The impact of the distance resolution on the error
in the effective volume should then be less severe.

To assess the error in the effective volume due to distance
resolution, we have computed the effective area for the same
exponential disk model and \emph{Gaia}-like RC selection as above at
all of the distance moduli of the combined extinction map of the
\appendixname. \figurename~\ref{fig:gaia_dist} shows the power
spectrum---calculated using a standard periodogram estimate---of the
effective area as a function of distance modulus (this includes an
extra factor of distance because of the transformation between
distance and distance modulus). The power spectrum at a given
resolution is the approximate error due to the ignorance of
sub-resolution fluctuations in the area; this approximation works well
because the power spectrum steeply declines with increasing $k$. An
estimate of the error in computing the effective volume using
Simpson's rule agrees with the power spectrum estimate. The black
curve in \figurename~\ref{fig:gaia_dist} displays the power spectrum
using the full-sky effective area. This leads to an error on the
$0.5\magunit$ spatial sampling of $\approx10^{-3}$. An investigation
of the angular dependence of the integrand reveals that much of this
error comes from the inner region of the MW, where both the extinction
and the density of an exponential disk are high and small-scale
structure in the extinction has a significant impact on the effective
volume. Excluding the inner regions (the red curve in
\figurename~\ref{fig:gaia_dist}) removes much of the error and leads
to an error of $\approx10^{-5}$. The error also decreases about twice
as fast with increasing resolution outside of the inner
regions. Increasing the distance sampling above that of
\citet{Green15a} by a factor of ten, which can be achieved within a
few years using the precise \emph{Gaia} parallaxes for many
stellar-extinction tracers, leads to errors of $\approx10^{-6}$ (full
sky) and $\approx10^{-10}$ (excluding the central regions). The latter
is good enough to allow sensitive measurements of the disk's structure
and asymmetries using the full \emph{Gaia} catalog, but studying the
inner regions with large \emph{Gaia} samples will be challenging.

The discussion so far has focused on angular and distance
\emph{precision}. However, \emph{accuracy} is as important for
disentangling intrinsic structural properties of the Milky Way's
stellar distribution from those of the dust distribution. In
\figurename~\ref{fig:ah-plane} below, we compare the distribution of
near-infrared (NIR) extinction in a few lines of sight for four
existing 3D extinction maps. While overall the agreement is good,
differences between different maps are much larger than the precision
of each of the maps. In particular, extinction maps constructed
largely using optical data (\citealt{Sale14a} and \citealt{Green15a})
disagree with IR-based maps in regions of high extinction. Because the
high-extinction regime is the most important for the error in the
effective volume, future extinction maps will need to incorporate IR
data to properly sample this regime.

\section{The effective survey volume for pencil-beam surveys}\label{sec:pencil}

\begin{figure*}[t!]
\begin{center}
\includegraphics[width=0.455\textwidth]{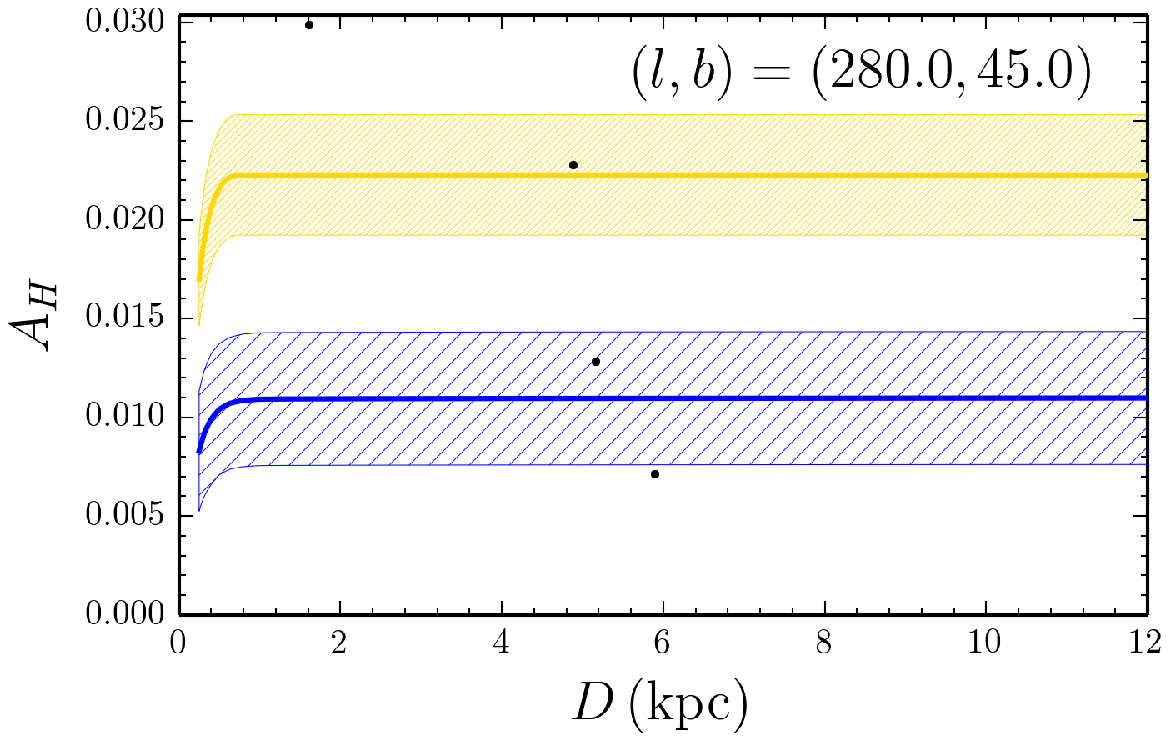} 
\includegraphics[width=0.455\textwidth]{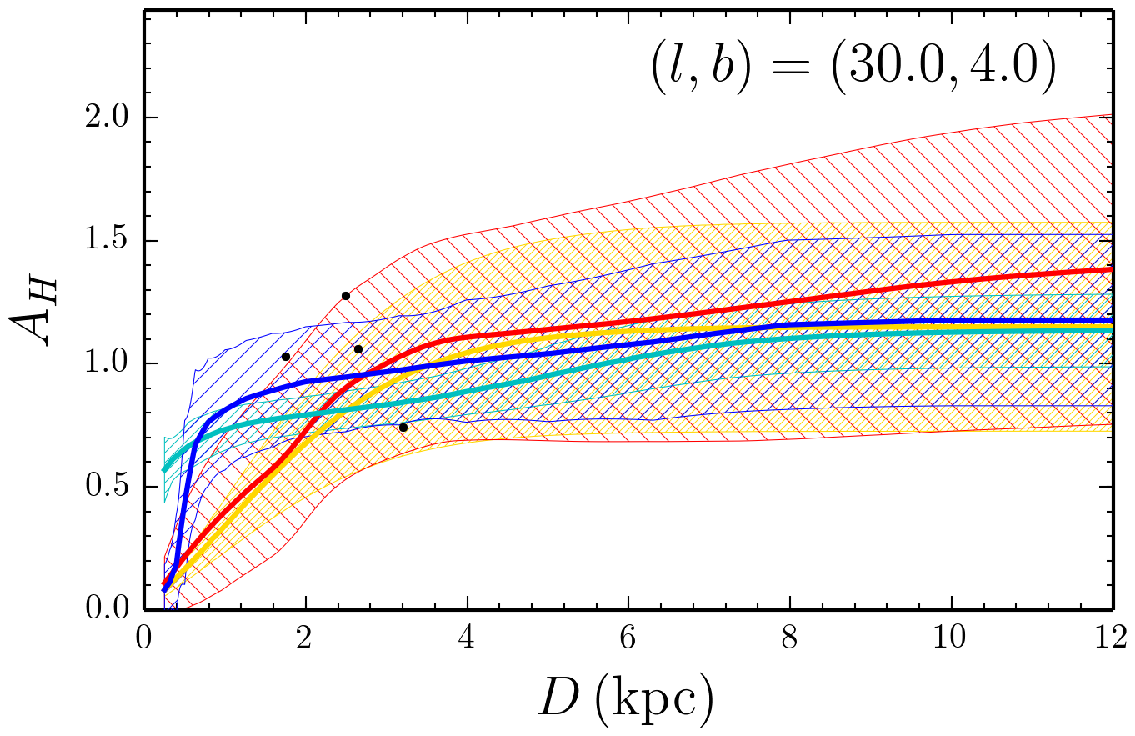}\\ 
\includegraphics[width=0.455\textwidth]{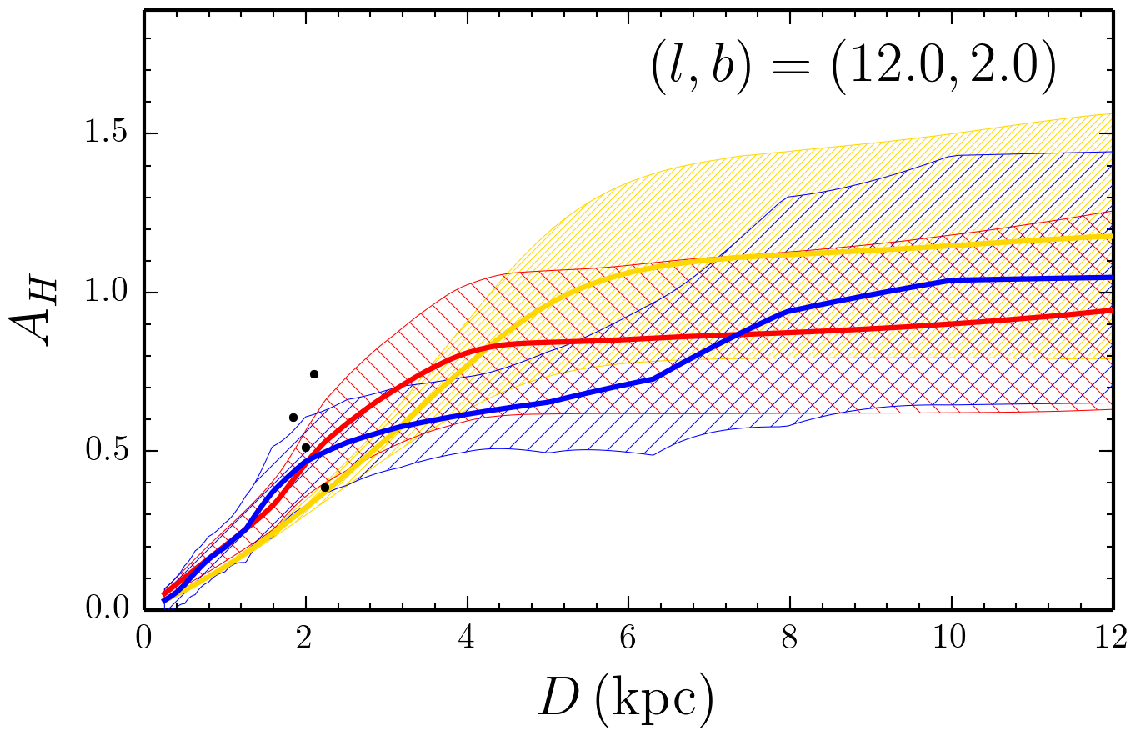} 
\includegraphics[width=0.455\textwidth]{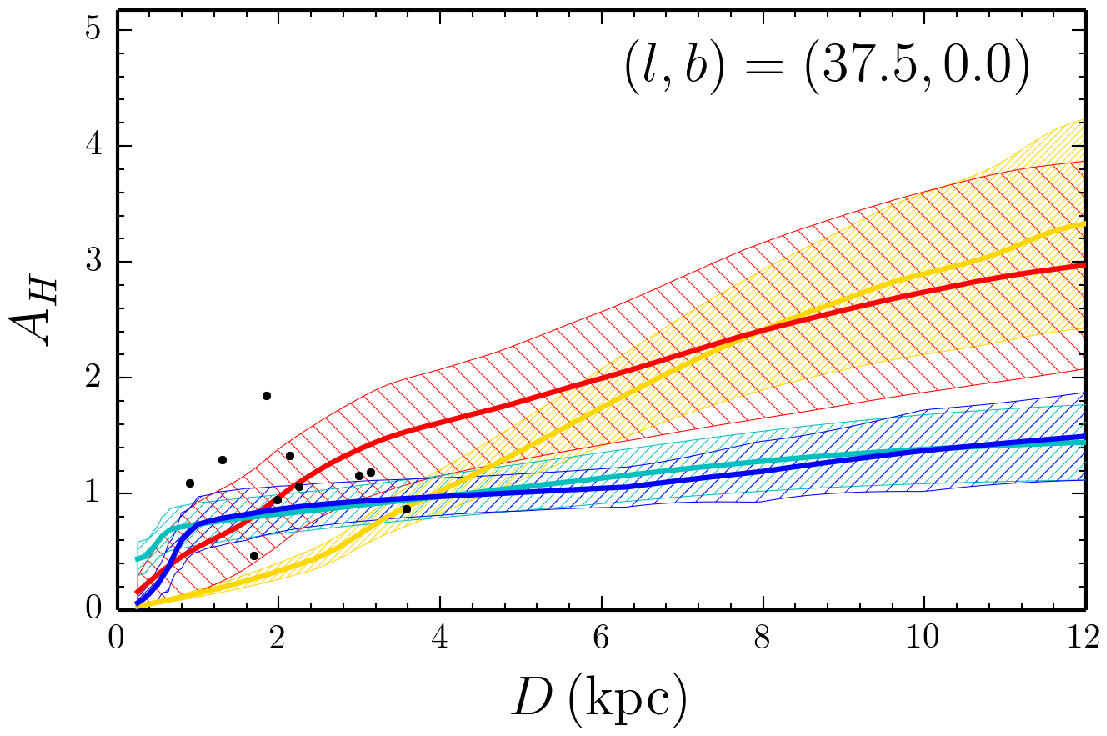}\\ 
\includegraphics[width=0.455\textwidth]{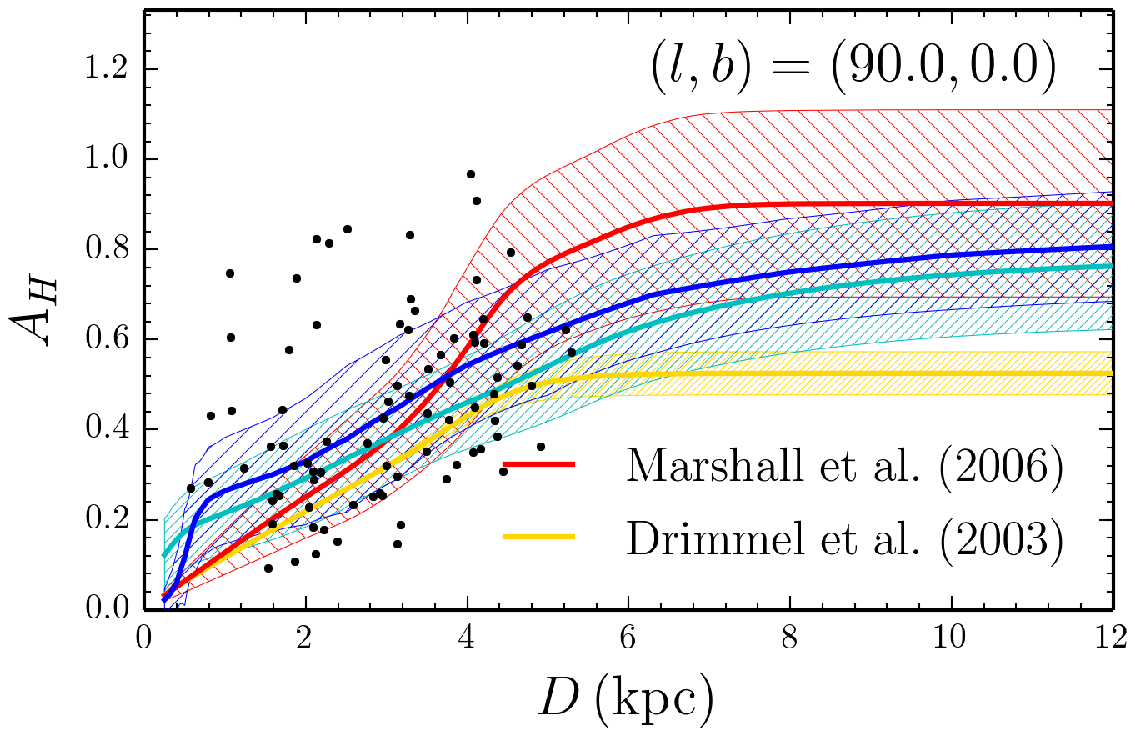} 
\includegraphics[width=0.455\textwidth]{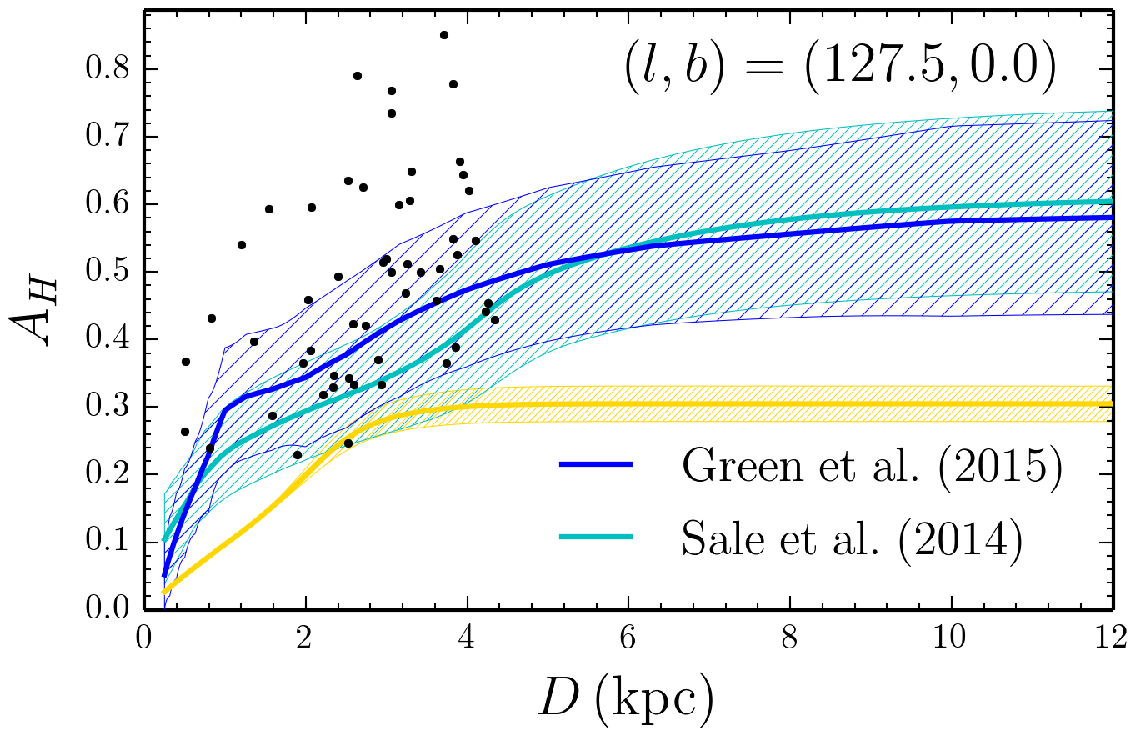}\\ 
\includegraphics[width=0.455\textwidth]{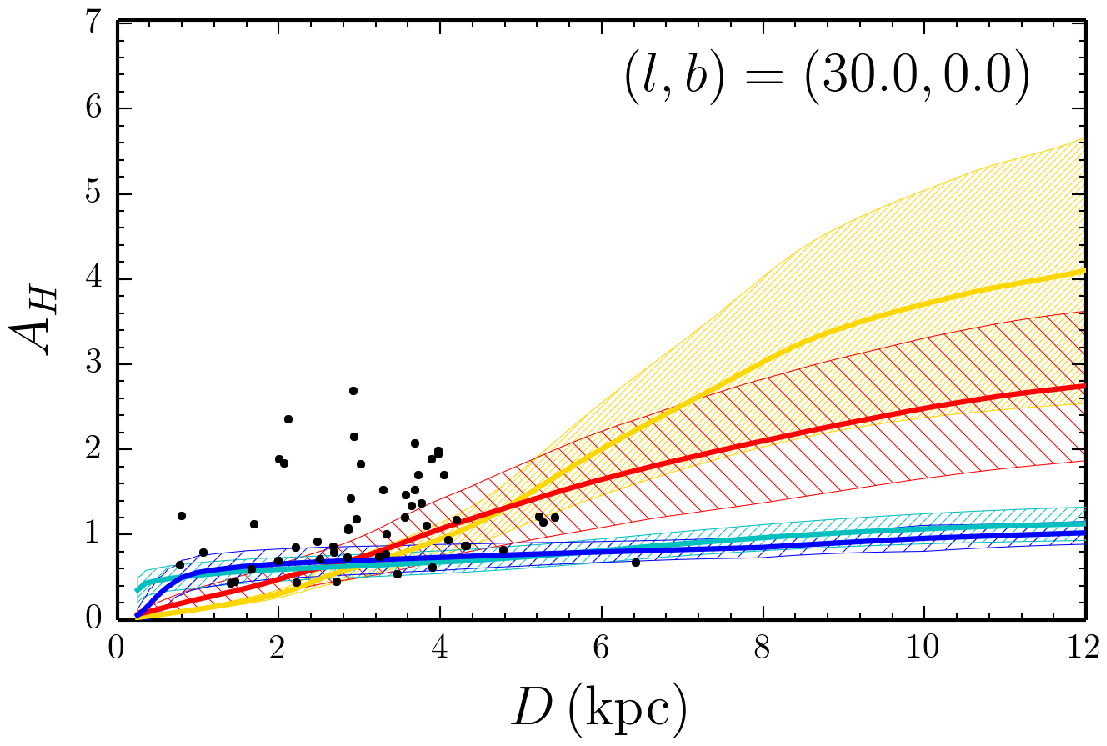} 
\includegraphics[width=0.455\textwidth]{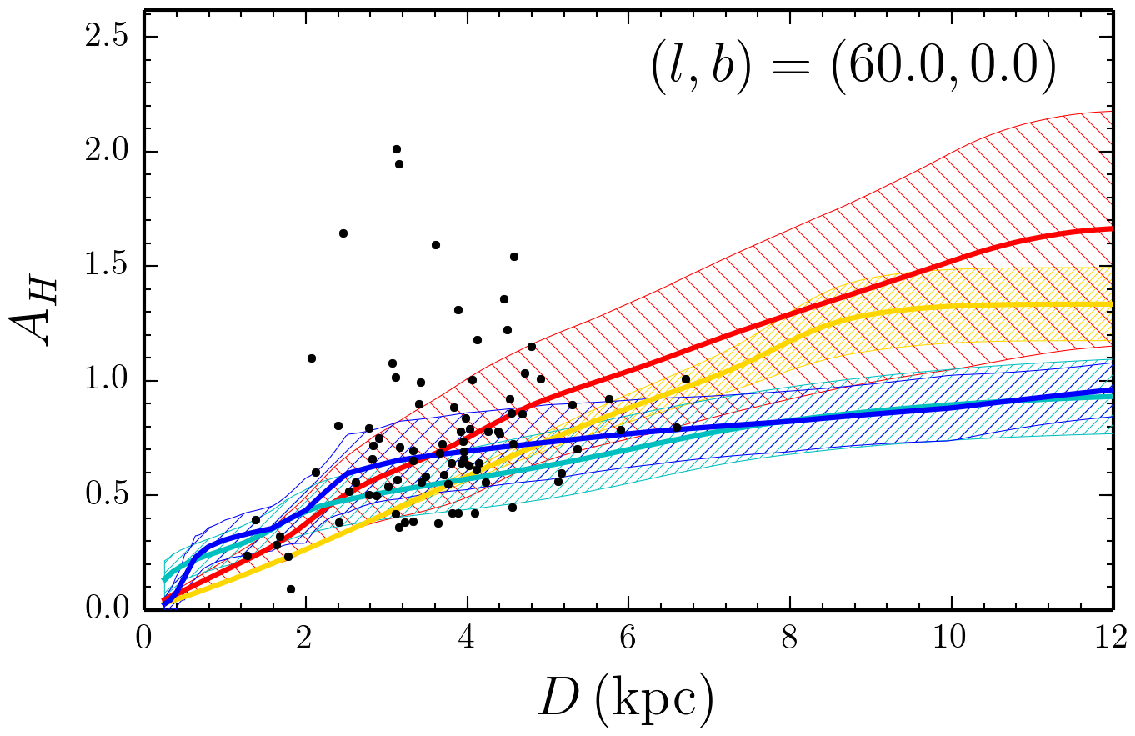} 
\end{center}
\caption{Distribution of extinction $A_H$ as a function of distance
  for four different three-dimensional extinction maps and eight
  representative APOGEE pointings. The solid line for each map
  displays the mean extinction over the area of an APOGEE pointing
  (which have radii of $1.49^\circ$, except for the $(l,b) =
  (12.0,2.0)$ pointing that has a radius of $1^\circ$) and hatched
  regions demonstrate the $1$ sigma range of the distribution. The
  dots are data points from the APOGEE-RC catalog. Overall the
  different extinction maps agree well with each other and with the
  data (the large discrepancies between the
  \citet{Sale14a}/\citet{Green15a} and
  \citet{Drimmel03a}/\citet{Marshall06a} maps are beyond where the
  former maps are considered reliable).}\label{fig:ah-plane}
\end{figure*}

\subsection{Formalism}\label{sec:pencil-formalism}

In principle the same formalism as presented above could be applied to
a pencil-beam survey. However, the sharp edges of each celestial
pointing in pencil-beam surveys would introduce significant ringing in
the effective selection function at each order, while apodizing the
boundary of each field would require much data to be
discarded. Pencil-beam surveys are closer to the case of a set of
delta-function pointings on the sky, where we would compute the
effective volume by summing over these pointings.

As an example of a pencil-beam survey we will consider the selection
function of RC stars in the APOGEE survey. For this survey the survey
selection function is accurately known \citep{BovyRC} and it is a
function of the field and apparent $H$ magnitude (not corrected for
extinction). That is, $S \equiv S(\field,H)$. The RC is close to a
standard candle in the NIR, with variations as a
function of color and metallicity of
$\approx0.1\magunit$. Nevertheless, we will continue to present the
formalism for the general case where there are variations of the
absolute magnitude with color $c$ (typically $c\equiv \jk$ for the NIR
RC) and \feh. With APOGEE in mind, we use $H$ instead of the generic
magnitude $m$ in this section and we again replace the arguments
$(l,b,D,c,\feh)$ of $H$ by $[\cdot]$ in many expressions to avoid
notational clutter.

The effective volume for a pencil-beam survey can be written as a sum
over field locations
\begin{widetext}
\begin{equation}\label{eq:effvolsimple}
\begin{split}
  \int \dd O \, \lambda(O|\theta) = \sum_{\mathrm{\field s}} \int &
  \dd D\,D^2\,\dens([X,Y,Z](D,\field)|\theta)\,\times\,\iint \dd
  c\,\dd\feh\, \rho(c,\feh)\, \times\iint \dd l\,\dd b\,
  \cos b\, S(\field,H[\cdot])\,,
\end{split}
\end{equation}
\end{widetext}
where we have assumed that the density $\dens(l,b,D)$ is sufficiently
smooth that it can be considered constant over the extent of the
field. When this is not the case, $\dens(l,b,D)$ could be expanded in
terms of basis functions around the center of each field in much the
same way as in \equationname~(\ref{eq:dens_ylm}), which would
introduce higher-order effective selection functions below, in the
same manner as in \equationname~(\ref{eq:effsel_cont}). We further
assume that the absolute-magnitude spread of the stellar tracer does
not have significant variations within a field; field-to-field
variations could be included, but are also neglected here.

From the expression in \equationname s~(\ref{eq:effvolsimple}), it is
clear that the three-dimensional distribution of extinction only
enters in the integration over $(l,b)$. We then again introduce the
\emph{effective selection function} \essf. For a standard
candle, \essf\ is the selection function averaged over $(l,b)$
\refstepcounter{equation}\begin{align}\begin{split} \essf(\field,D) &
    = \iint \frac{\dd l\,\dd b}{\Omega}\,\cos
    b\,S(\field,H[l,b,D])\,,\notag\\&\pushright{(\mathrm{standard\ candle};\theequation)}
\end{split}\end{align}
where $\Omega$ is the area of the $(l,b)$ integration. If the tracer
stellar population is not a standard candle, we absorb the
integration over $(c,\feh)$ into the definition of the effective
selection function
\refstepcounter{equation}\begin{align}\label{eq:effsel-apogee}
\begin{split}
  & \essf(\field,D) = \,\iint \dd c\,\dd\feh\, \rho(c,\feh)\\
  & \,\times\iint \frac{\dd l\,\dd b}{\Omega}\,\cos b\,S(\field,H[l,b,D,c,\feh])\,;\notag\\&\pushright{(\mathrm{general\ case};\theequation)}
\end{split}
\end{align}
This integration over $(c,\feh)$ can again be easily performed using
Monte Carlo integration.

With this definition of the effective selection function, the
effective volume becomes simply
\begin{align}
\begin{split}
  & \int \dd O \, \lambda(O|\theta) =\\& \quad  \sum_{\mathrm{\field s}} \Omega\,\int \dd D\,D^2\,\dens([X,Y,Z](D,\field)|\theta)\,\essf(\field,D)\,.
\end{split}
\end{align}

\subsection{The effective selection function of the APOGEE-RC sample}\label{sec:pencil-example}

The APOGEE selection function is a constant $S(\field,k)$ within each
bin $k$ of a small number of magnitude bins
$[H_{\mathrm{min},k},H_{\mathrm{max},k}]$
\citep{Zasowski13a,BovyRC}. Therefore, we can simplify \essf\ to
\begin{widetext}
\begin{equation}
  \essf(\field,D) = \sum_k S(\field,k)\,\frac{\Omega(H_{\mathrm{min},k}-H_0(D) < A_H(l,b,D) < H_{\mathrm{max},k}-H_0(D))}{\Omega_f}\,,
\end{equation}
\end{widetext}
for a standard candle with $H_0 = M_H + \mu$, where $M_H$ is the
absolute magnitude of the standard candle and $\mu$ is the distance
modulus; $\Omega(H_{\mathrm{min},k}-H_0 < A_H(D) <
H_{\mathrm{max},k}-H_0)$ is the area of the field with $A_H$ between
the given boundaries and $\Omega_f$ is the total area of the field
(not all APOGEE fields have the same area). For a non-standard candle
we additionally integrate over $(c,\feh)$, which in the APOGEE-RC case
are $(\jksq,\feh)$. In this expression, we have suppressed the
dependence of $A_H$ on $(l,b)$ for clarity. The APOGEE selection
function $S$ and effective selection function \essf\ for a standard
candle or any tracer with a known $M_H$ distribution is implemented as
part of the \texttt{apogee} Python package
at\\ \centerline{\url{http://github.com/jobovy/apogee}} as part of its
\texttt{apogee.select.apogeeSelect} module.

Thus, at any given distance the effective selection function only
depends on a smoothed version of the two-dimensional extinction
distribution at that distance: the fractional area of the field in
wide ($\approx 1 \magunit$) bins in $A_H(D)$. For the RC sample that
we use here the dependence of $H_0$ on $(\jksq,\feh)$ is weak
($\lesssim 0.1\magunit$; see \figurename~3 of \citealt{BovyRC}), such
that we can treat the RC as a standard candle. The weak dependence of
the RC absolute magnitude implies that any density or phase-space fits
are only very weakly dependent on the color-metallicity distribution
of stars in the sample.

\begin{figure*}[t!]
\includegraphics[width=0.48\textwidth,clip=]{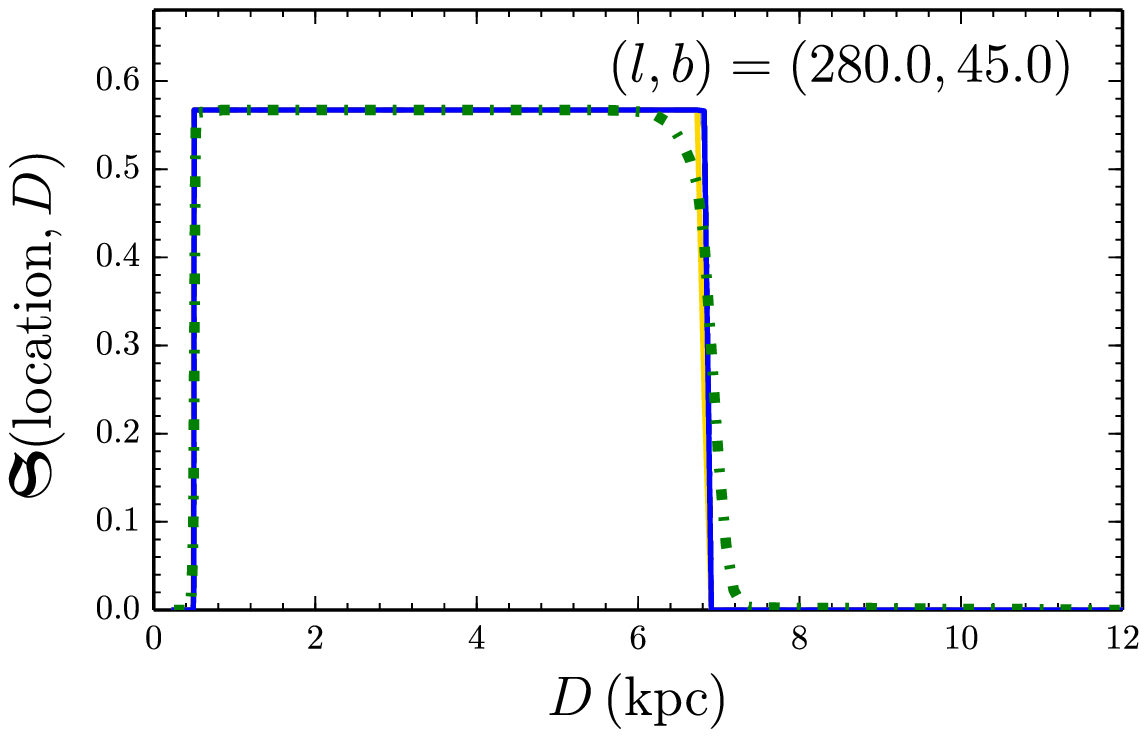} 
\includegraphics[width=0.48\textwidth,clip=]{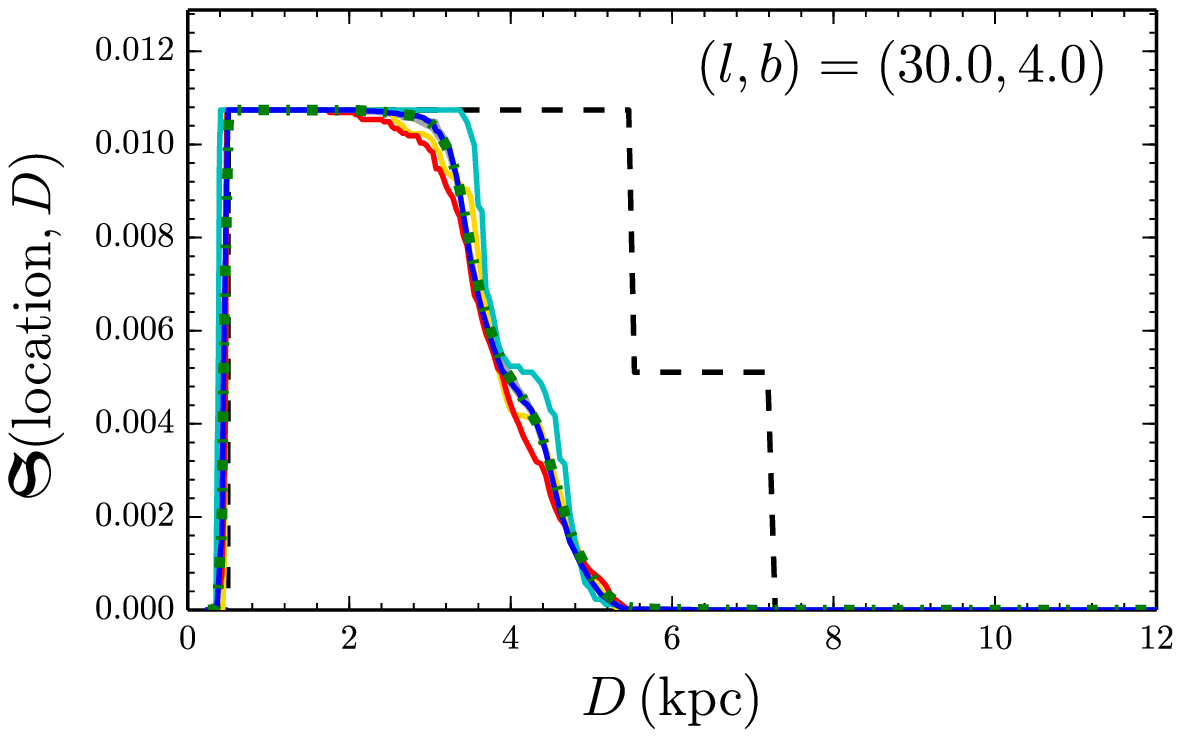}\\ 
\includegraphics[width=0.48\textwidth,clip=]{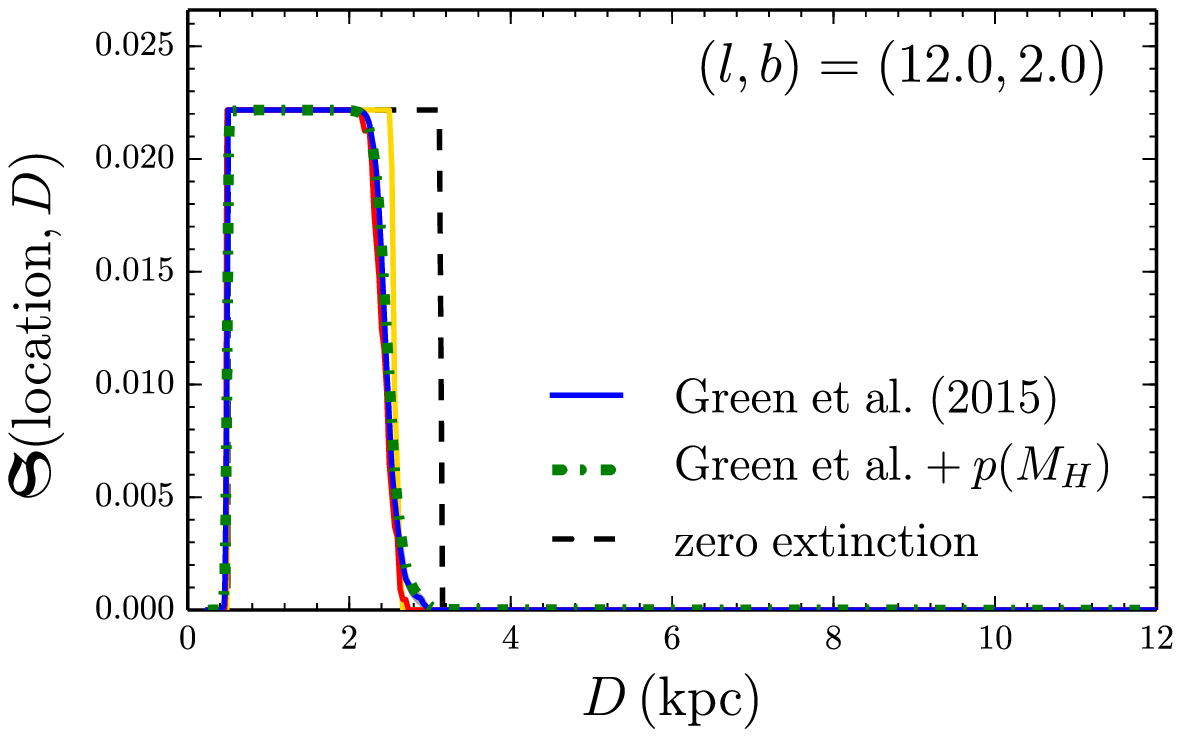} 
\includegraphics[width=0.48\textwidth,clip=]{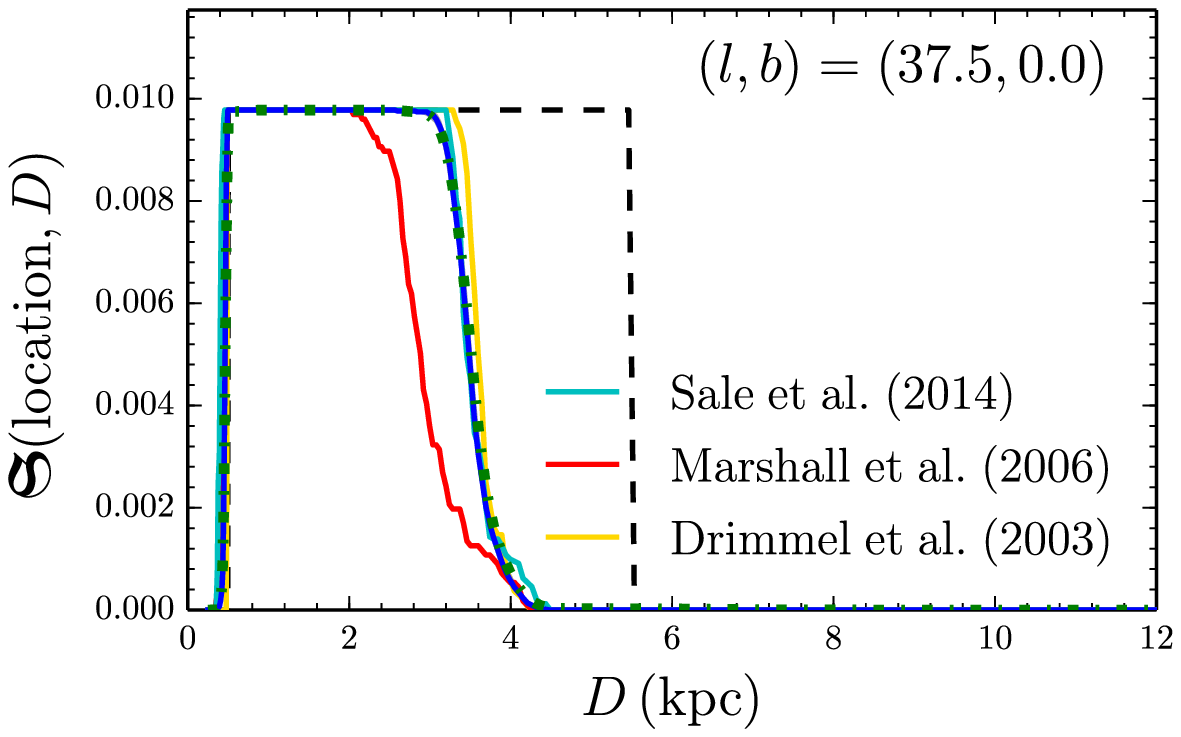}\\ 
\includegraphics[width=0.48\textwidth,clip=]{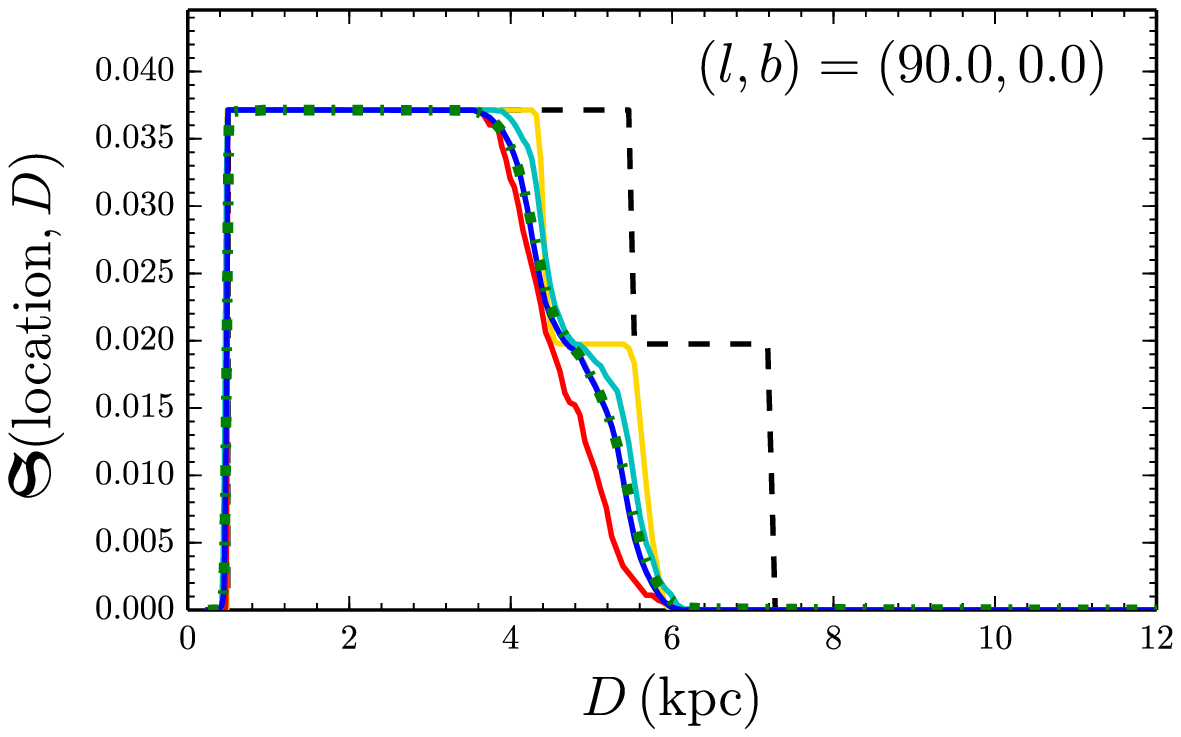} 
\includegraphics[width=0.48\textwidth,clip=]{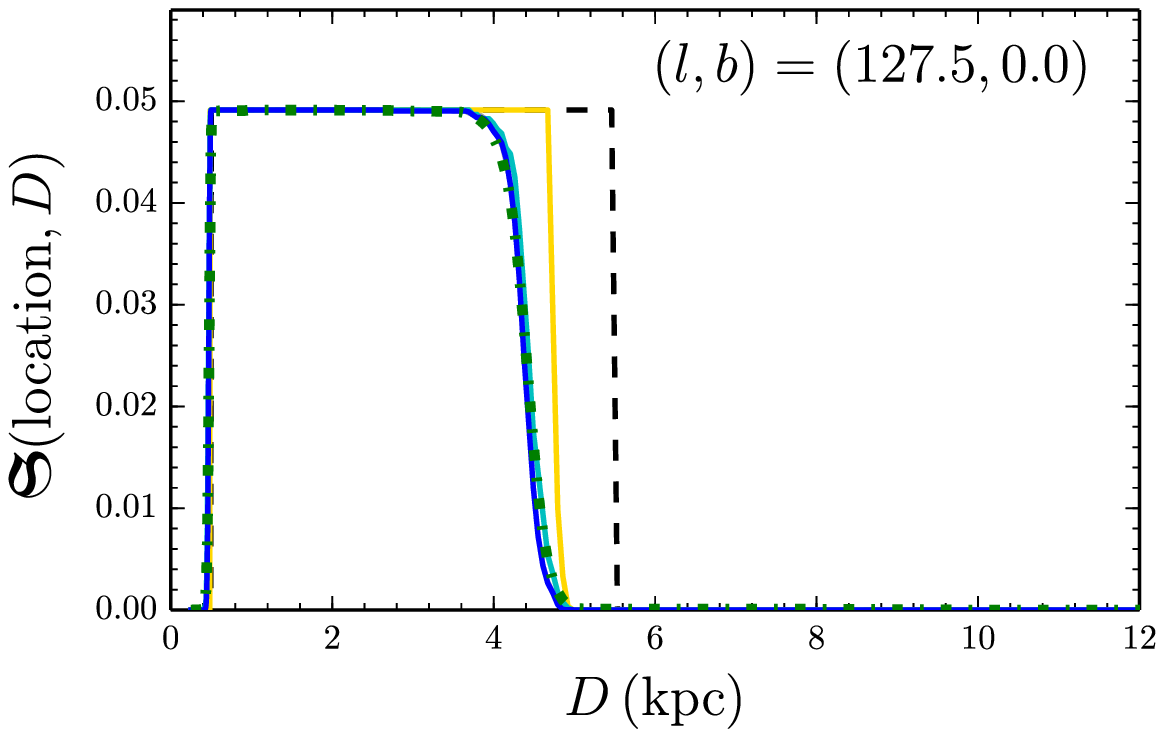}\\ 
\includegraphics[width=0.48\textwidth,clip=]{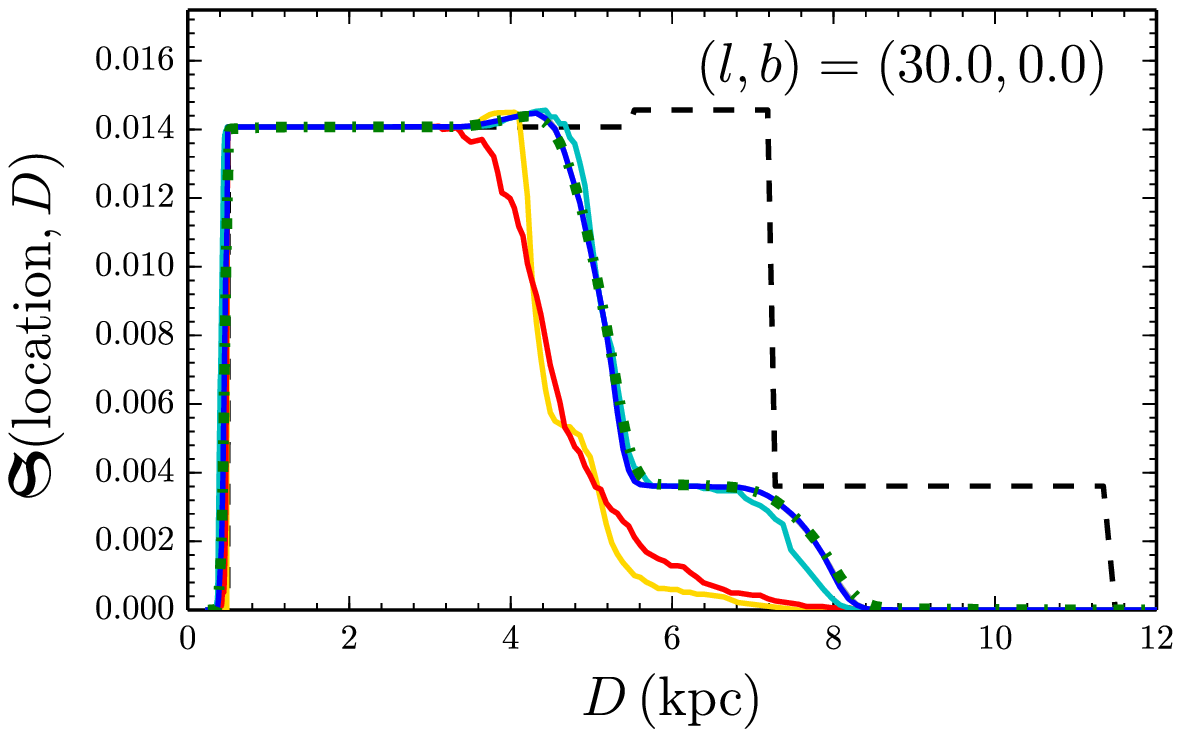} 
\includegraphics[width=0.48\textwidth,clip=]{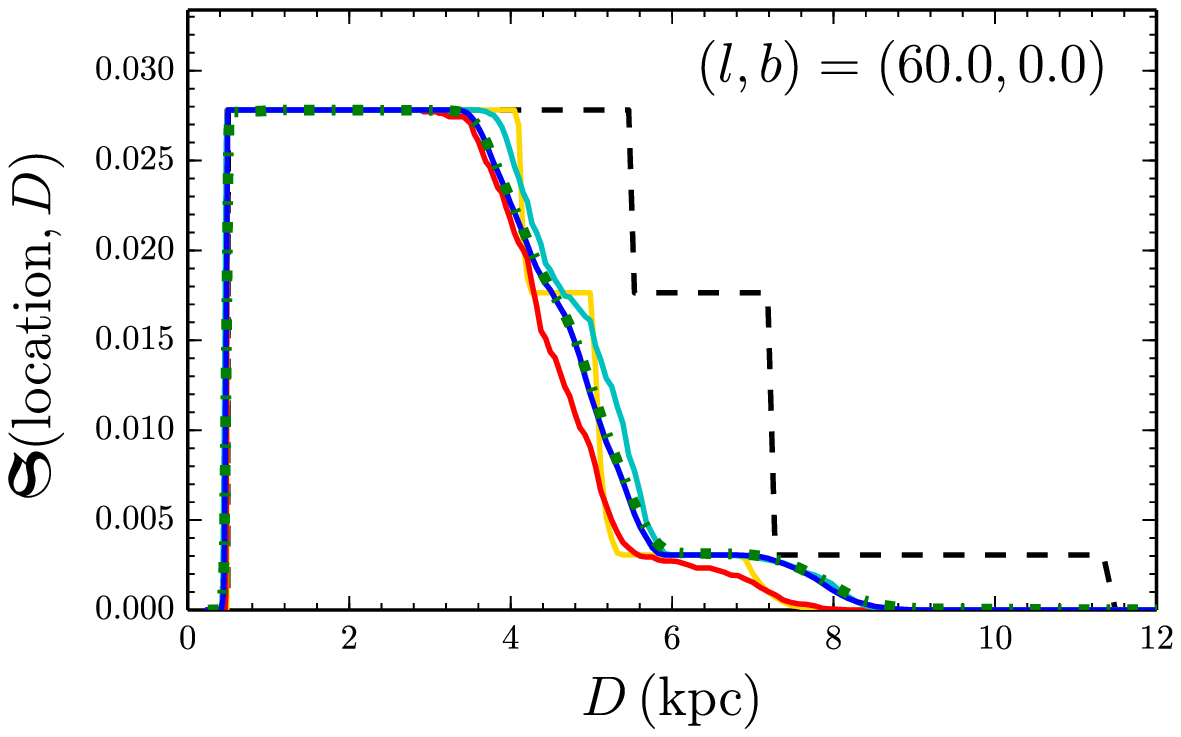} 
\caption{Effective selection function $\essf(\mathrm{location},D)$ for
  eight APOGEE pointings. The dashed line displays the raw selection
  function, which is a piecewise-constant function over one, two, or
  three magnitude ranges, translated from $H$ to distance assuming
  $M_H=-1.49$ for the RC. The blue solid line shows the effective
  selection function calculated using the \citet{Green15a} 3D extinction map
  assuming that the RC is a standard candle, while the dot-dashed
  green line takes into account the spread in $M_H$ in the RC for the
  whole APOGEE-RC sample. The uncertainty in this effective selection
  function as calculated from the \citet{Green15a} MCMC samples is
  smaller than the width of the line. The effective selection function
  for three other extinction maps is also shown where these maps are
  available (\citealt{Drimmel03a}: full sky; \citealt{Marshall06a};
  $-100^\circ \leq l \leq 100^\circ$ and $|b| \leq 10^\circ$;
  \citealt{Sale14a}: $30^\circ \leq l \leq 215^\circ$ and $|b| \leq
  5^\circ$). The \citet{Marshall06a} map generally leads to a
  shallower effective selection function, because it has on average
  higher extinction (see \figurename~\ref{fig:ah-plane}). The
  agreement between the two most recent extinction maps (\citealt{Sale14a}
  and \citealt{Green15a}) is remarkably good.}\label{fig:effsel-plane}
\end{figure*}

To illustrate the pattern of extinction that affects APOGEE
observations, we can interpret the extinction displayed in
\figurename~\ref{fig:dust} as the $H$-band extinction between $0$ and
$1.1$ ($A_G / A_H \approx5.1$). This upper limit corresponds to the
maximum extinction for which we can see the RC to $\approx4.5\kpc$ in
a medium-deep APOGEE pointing ($H < 12.8$). This gives a sense of the
volume over which the RC can be mapped with APOGEE. RC stars can be
seen to $4\kpc$ over almost the whole sky, while at $5\kpc$ and
further much of the disk region is obscured. APOGEE has deep pointings
(down to $H < 13.8$) along a few lines-of-sight that allow for some of
this highly-extinguished structure to be penetrated.

In \figurename~\ref{fig:ah-plane} we demonstrate the range of
extinction values in eight APOGEE pointings for four different
three-dimensional extinction maps. The default extinction model that
we use is that of \citet{Green15a}, which is displayed in blue. Each
curve shows the mean extinction within the APOGEE pointing and the
hatched regions display the $1\sigma$ range of the extinction
distribution. Other three-dimensional extinction maps are compared to
our default model: the yellow curve displays the extinction model of
\citet{Drimmel03a}, which is a fully-sky map with an angular
resolution of $\approx20'$; the red curve shows the 3D map of
\citet{Marshall06a} which is available over $-100^\circ \leq l \leq
100^\circ$ and $|b| \leq 10^\circ$; and the cyan curve gives the map
of \citealt{Sale14a}, available over $30^\circ \leq l \leq 215^\circ$
and $|b| \leq 5^\circ$. The agreement between the recent extinction
maps of \citet{Sale14a} and \citet{Green15a} is remarkably good, but
they both typically give smaller extinction close to the Galactic
plane than the \citet{Marshall06a} map. The APOGEE-RC data themselves
provide good estimates of extinction values at the range of distances
included in our sample and our data are displayed as black
dots. Overall the agreement between the extinction maps and our data
is good; however, our extinction vs. distance data do not allow us to
unambiguously prefer the \citet{Marshall06a} map over the
\citet{Green15a} map at low Galactic latitudes.

While the differences between the different 3D extinction maps can be
quite substantial, especially at large distances, they only matter for
any density inference in as much as they lead to different effective
selection functions
$\essf(\field,D)$. \figurename~\ref{fig:effsel-plane} demonstrates the
effective selection function for the same eight pointings as displayed
in \figurename~\ref{fig:ah-plane}. At Galactic latitudes above a few
degrees, the different maps lead to similar effective selection
functions, even though the extinction can be substantial ($A_H \approx
1$ near the faint end of the survey). We have included one APOGEE
pointing in the inner Galaxy at $(l,b) = (12,2)$: these locations were
only observed down to $H < 11$ and therefore the effective selection
function only reaches to $\approx3\kpc$; the different extinction maps
all return the same $\essf$ for such inner-Galaxy pointings. Within a
few degrees of the mid-plane, larger differences occur between the
effective selection function calculated using the \citet{Marshall06a}
map and the \citet{Green15a}/\citet{Sale14a} maps, especially at $l
\lesssim60^\circ$.

All but one of the curves in \figurename~\ref{fig:effsel-plane} are
computed assuming that the RC is a standard candle. The dashed green
line demonstrates what happens when we relax this assumption for the
\citet{Green15a} map. For all but the least obscured pointings, the
small $\approx0.1\magunit$ width of the RC absolute magnitude has no
effect on the effective selection function, essentially because the
small-scale structure in the extinction map washes out this small
spread. Only at high Galactic latitude is the extinction small enough
that the small $M_H$ spread in the RC can affect the effective
selection function.

We estimate the error in the calculation of the effective volume due
to the distance integration in the same manner as for the \emph{Gaia}
example above for the eight fields of \figurename s~\ref{fig:ah-plane}
and \ref{fig:effsel-plane}. We find that the current error is
$\approx\mathrm{few}\times10^{-3}$, decreasing by $\approx10^{-2}$ per
decade of improved distance resolution. These errors are larger than
for the \emph{Gaia} example above, mainly because of the much narrower
magnitude bins employed by APOGEE and due to APOGEE's focus on the
mid-plane, where the extinction has more small-scale
structure. Analyses using hundreds of tracers are therefore unaffected
by the current distance distance sampling. Using larger samples may
lead to systematic uncertainties due to the limited distance sampling
in current extinction maps, but again the relevant error is the
relative error between models that fit the data similarly well
($\Delta \chi^2 = \mathcal{O}(1)$) and this is likely much smaller
than the absolute error discussed here. Mock-data tests for an
APOGEE-like sample of 20,000 stars by \citet{Bovy15a} indicate that
such samples are also currently unaffected by the distance
resolution. As discussed above, improvements in the distance
resolution by about a factor of 10 are likely in the near future, such
that spectroscopic samples close to the mid-plane of $\approx10^6$
stars should be able to be analyzed; for spectroscopic surveys that
avoid the mid-plane (\ie, optical surveys) the requirements are less
severe.

\section{Conclusions}\label{sec:discussion}

Motivated by the advent of near-infrared spectroscopic surveys and
\emph{Gaia}, we have discussed likelihood-based phase-space inference
in the MW in detail in this paper, focusing in particular on the
effects of interstellar extinction. Central to this inference is the
effective survey volume, which acts as a normalization factor in the
likelihood. In principle the computation of this effective volume is
straightforward: one simply integrates the density at $(l,b,D)$
multiplied by the selection function, which includes the effects of
extinction, over the full MW volume. In practice, this integration is
expensive and the uncertainties due the angular and distance
resolution in the extinction map have been difficult to assess.

In \sectionname~\ref{sec:large}, we have introduced a novel formalism
that introduces the \emph{effective selection function} to assess the
impact of angular and distance resolution on the computation of the
effective volume for a large-area survey such as \emph{Gaia}. We have
found that the angular resolution of current extinction maps is
sufficient to analyze \emph{Gaia} samples consisting of millions of
stars in terms of smooth, axisymmetric density distribution. Only
modest and realistic improvements in the angular resolution are
necessary to analyze the full \emph{Gaia} data set ($\approx10^9$
stars; in case one is so inclined). More limiting than the angular
resolution is the distance sampling of current maps. The current
distance sampling is about $0.5\magunit$, or about $25\,\%$ in
distance, which causes an uncertainty in the effective volume of
$\approx10^{-3}$ to $10^{-5}$, depending on how much of the inner MW
is included. From an analysis of the small-scale structure in the
distance integration, we expect this uncertainty to decrease by three
orders of magnitude or five (excluding the inner MW) for an
improvement in the distance resolution by a factor of ten. Such an
improvement appears plausible with the precise \emph{Gaia} distances
for large numbers of stellar tracers of the extinction and allows
sensitive analyses of the disk's structure with the full \emph{Gaia}
data set. Analyzing the non-axisymmetric component of the stellar disk
that is less smooth than the exponential disk models considered here
would place somewhat more stringent conditions on the angular and
distance resolution of extintion maps; these can be evaluated with a
similar formalism as introduced here. Extinction maps also need to
attain a higher accuracy, because current extinction maps differ by
much more than their stated uncertainties in regions of high
extinction. This requires a larger contribution from IR data in the
creation of 3D extinction maps.

We have also discussed the formalism for phase-space inference of
non-contiguous, spectroscopic surveys. An application of this
formalism is given in \citet{Bovy15a}, who analyze the density
distribution of abundance-selected populations using RC stars in
APOGEE. We have demonstrated that the 3D extinction map can be fully
and efficiently taken into account in likelihood-based phase-space
inferences using an effective selection function under the assumption
that the density is close to constant within each spectroscopic
pointing.

Our formalism makes it clear that in both the contiguous and the
non-contiguous case the small-scale structure in the 3D extinction map
is filtered by the smooth density field. This is because the effective
survey volume is essentially the cross-correlation of the smooth
(phase-space) density and the effective selection function; as the
density is very smooth, this cross-correlation is effectively a
low-pass filtering of the effective selection function, which removes
the influence of the patchy extinction map.

Progress in obtaining better 3D extinction maps in the future will
require a balance between angular and distance resolution because of
the finite number of stars available to use as tracers. Essentially,
one can use larger angular bins to obtain a larger number of stars to
infer the distance dependence of the extinction, although in practice
a better approach is to require a high degree of correlation between
adjacent, small angular bins or by using Gaussian processes
\citep[\eg,][]{Sale14b}. From our analysis we can conclude that
improving the distance resolution---especially in the high-extinction,
inner Galaxy---is much more important for studies of Galactic
structure than improving the angular resolution. This is especially
important for future analyses that aim to take advantage of the full
\emph{Gaia} catalog to tease out small signals of, \eg, phase-space
asymmetries. Such analyses require the best smooth background models
and the ability to compute the background model to an accuracy of
$\approx10^{-9}$.

\acknowledgements It is a pleasure to thank the anonymous referee and
Wilma Trick for helpful comments. Some of the results in this paper
have been derived using the HEALPix \citep{Gorski05a} and
\texttt{healpy} packages. J.B. received support from a John N. Bahcall
Fellowship, the W.M. Keck Foundation, and the Natural Sciences and
Engineering Research Council of Canada. H.W.R. received funding for
this research from the European Research Council under the European
Union's Seventh Framework Programme (FP 7) ERC Grant Agreement
n. [321035]. J.B. and H.W.R. acknowledge the generous support and
hospitality of the Kavli Institute for Theoretical Physics in Santa
Barbara during the `Galactic Archaeology and Precision Stellar
Astrophysics' program, where some of this research was performed.

\appendix
\section{A three-dimensional extinction map over the full sky}\label{sec:appendix}

In recent years a number of three-dimensional maps of the integrated
extinction out to many kpc based on modeling stellar photometry have
appeared \citep[\eg,][]{Marshall06a,Sale14a,Green15a}. None of these
cover the full sky. To illustrate the formalism for accounting for the
effect of extinction on the volume selection of stellar surveys
presented in this paper and to make projections for \emph{Gaia}, we
require a full-sky, three-dimensional extinction map.

We perform a simple combination of the extinction maps of
\citet{Marshall06a} (based on 2MASS data) and \citet{Green15a} (based
on Pan-STARRS and 2MASS data) and fill in the remaining parts of the
sky with the map of \citet{Drimmel03a}, a three component analytic
model for the dust distribution fit to the COBE DIRBE data and
normalized to the map of \citet{Schlegel98a}. \citet{Bovy15a} found by
modeling the observed counts of RC stars near the midplane in APOGEE
that the map of \citet{Marshall06a} performs better than that of
\citet{Green15a} where they overlap, essentially because the latter
underestimates the amount of extinction in regions of high extinction
due to the paucity of main-sequence stars that they primarily use,
while \citet{Marshall06a} mainly employ giant stars that are visible
to larger distances. We do not use the map of \citet{Sale14a}, as it
led to inferior fits of the APOGEE stellar densities.

Specifically, we follow the hierarchical HEALPix format of
\citet{Green15a} and evaluate the \citet{Marshall06a} map, which
exists over the range $-100^\circ \leq l \leq 100^\circ$ and $|b| \leq
10^\circ$, on a grid of HEALPix pixel centers with $N_{\mathrm{side}}
= 512$ or an approximate resolution of $7'$. We employ this resolution
to slightly oversample the native resolution of $15'$. We then add the
hierarchical map of \citet{Green15a}, which has a variable angular
resolution of $\approx4'$ to $14'$, outside of the \citet{Marshall06a}
area. The remaining area of the sky, mainly near the south celestial
pole, is filled in with the map of \citet{Drimmel03a} at a HEALPix
resolution of $N_{\mathrm{side}}= 256$, which again somewhat
oversamples the resolution of the \citet{Drimmel03a} map ($\approx
20'$). We also fill in parts of $N_{\mathrm{side}}= 256$ pixels that are
partially covered by the other two maps.

The resulting map of $A_G$ at $5\kpc$ is displayed in
\figurename~\ref{fig:dust}. We use $A_G / E(B-V) = 2.35$
\citep{BovyRC}. While this map is not perfect and the effect of
stitching together the three different extinction maps is clearly
visible, this map does an adequate job of describing the extinction
over the full sky and is good enough for our purposes here. The most
problematic region is the region of the Galactic plane covered by
neither the \citet{Marshall06a} or the \citet{Green15a} map
($250^\circ \lesssim l \lesssim 260^\circ$), but in the high-latitude
regions filling in uncovered regions with the model of
\citet{Drimmel03a} performs well.

To aid in the comparison of different 3D extinction maps we have made
a code publicly available
at\\ \centerline{\url{http://github.com/jobovy/mwdust}} This code
downloads and installs all of the necessary data and code to evaluate
the extinction maps of \citet{Schlegel98a}, \citet{Drimmel03a},
\citet{Marshall06a}, \citet{Sale14a}, and \citet{Green15a} using a
common interface.

\end{document}